\documentclass[12pt]{iopart}
\pdfoutput=1
\expandafter\let\csname equation*\endcsname\relax
\expandafter\let\csname endequation*\endcsname\relax
\usepackage{cite}
\usepackage{subfig}
\usepackage{subcaption}
\usepackage{amsmath}
\usepackage[protrusion=true,expansion=true]{microtype}
\usepackage{times}
\usepackage{graphicx,bm,xcolor,braket, tikz}
\usepackage[normalem]{ulem}
\usepackage{hyperref}
\hypersetup{colorlinks,linkcolor={blue},citecolor={blue},urlcolor={blue}}
\usepackage{amssymb}
\usepackage[utf8]{inputenc}
\usepackage[english]{babel}

\makeatletter
\def\graphicscale{\twocolumn@sw{0.3}{0.4}}
\def\graphicthreescale{\twocolumn@sw{0.3}{0.4}}

\newcommand{\be}{\begin{equation}}
\newcommand{\ee}{\end{equation}}
\newcommand{\bea}{\begin{align}}
\newcommand{\eea}{\end{align}}

\usepackage{amsmath}

\usepackage{dsfont}
\usepackage{mathrsfs}
\usepackage{fancyhdr}
\usepackage{esint}
\DeclareMathOperator{\sign}{sign}
\usepackage{comment}
\renewcommand{\subsectionmark}[1]{}
\begin{document}
\title{Quasicondensation and off-diagonal long-range order of hard-core bosons during a free expansion}

\author{A. Takács~$^{1,*}$, S. Scopa~$^{2,3}$,  P. Calabrese~$^{2,4}$, L. Vidmar~$^{5,6}$ and J. Dubail~$^{1,7}$}

\noindent
\begin{center}{\small\it
$^1$~Universit\'e de Lorraine, CNRS, LPCT, F-54000 Nancy, France\\
$^2$~SISSA and INFN, via Bonomea 265, 34136 Trieste, Italy\\
$^3$~Laboratoire de Physique Th\'eorique et Mod\'elisation, CNRS UMR 8089,
	CY Cergy Paris Universit\'e, 95302 Cergy-Pontoise Cedex, France\\
$^4$~International Centre for Theoretical Physics (ICTP), Strada Costiera 11, 34151 Trieste, Italy\\
$^5$~Department of Theoretical Physics, J. Stefan Institute, SI-1000 Ljubljana, Slovenia\\
$^6$~Department of Physics, Faculty of Mathematics and Physics, University of Ljubljana, SI-1000 Ljubljana, Slovenia \\
$^7$~CESQ and ISIS (UMR 7006), University of Strasbourg and CNRS, 67000 Strasbourg, France}
\end{center}
{\small \href{attila.takacs@univ-lorraine.fr}{*attila.takacs@univ-lorraine.fr}}

\date{\today}

\begin{abstract}
Quasicondensation in one dimension is known to occur for equilibrium systems of hard-core bosons (HCBs) at zero temperature. This phenomenon arises due to the off-diagonal long-range order in the ground state, characterized by a power-law decay of the one-particle density matrix $g_1(x,y)\sim |x-y|^{-1/2}$~--~a well-known outcome of Luttinger liquid theory. 
Remarkably, HCBs, when allowed to freely expand from an initial product state (i.e., characterized by initial zero correlation), exhibit quasicondensation and demonstrate the emergence of off-diagonal long-range order during nonequilibrium dynamics. 
This phenomenon has been substantiated by numerical and experimental investigations in the early 2000s.
In this work, we revisit the dynamical quasicondensation of HCBs, providing a fully analytical treatment of the issue. 
In particular, we derive an exact asymptotic formula for the equal-time one-particle density matrix by borrowing ideas from the framework of quantum Generalized Hydrodynamics. 
Our findings elucidate the phenomenology of quasicondensation and of dynamical fermionization occurring at different stages of the time evolution, as well as the crossover between the two. 
\end{abstract}
\maketitle
\nopagebreak

\section{Introduction}\label{sec:intro}

The possibility to experimentally realize many-body quantum systems, which are brought far from equilibrium and evolve in nearly perfect isolation from environment~\cite{greiner_mandel_02b, kinoshita2006quantum}, has opened new challenges for a theoretical description of quantum dynamics. 
Exact solutions of thermodynamically large interacting systems are particularly desirable.
They may be achieved for certain classes of initial states of integrable quantum systems, however, even in these cases, obtaining a complete solution is far from straightforward~\cite{Calabrese_2016, Caux_2016, Essler_2016}.

An intriguing example of a quantum many-body system are interacting bosons in the strongly repulsive regime, known also as hard-core bosons (HCBs). In one dimension, due to the virtue of the Holstein-Primakoff and Jordan-Wigner transformation~\cite{cazalilla2011one}, they can be mapped into noninteracting spinless fermions, and consequently, one can design efficient numerical schemes to calculate their nonequilibrium properties. Nevertheless, analytical solutions and closed-form expressions for the time-dependence of certain observable  s have remained elusive, and almost restricted to numerical studies e.g.~\cite{rigol2004universal,girardeau2000breakdown,vignolo2013universal}.

The nontrivial property of one-dimensional (1D) HCBs is that, despite a mapping to noninteracting fermions, many of their nonequilibrium properties differ from those of fermions. 
For example, while the momentum distribution of translationally invariant fermions either does not evolve in time (for periodic boundaries) or it exhibits persistent oscillations (for open boundaries)~\cite{wright_rigol_14}, the momentum distribution function of HCBs equilibrates to a stationary value~\cite{rigol2007relaxation, rigol_muramatsu_06} that is described by the generalized Gibbs ensemble~\cite{Vidmar_2016}.
Moreover, in their ground state 1D HCBs display quasicondensation~\cite{rigol2004universal}.
A surprising observation is that characteristics of quasicondensation, particularly at nonzero momenta, may also manifest in far-from-equilibrium dynamics~\cite{Rigol2004quasicond}. 
The explanation of this phenomenon was offered in the context of the emergent eigenstate solution to quantum dynamics~\cite{vidmar_iyer_17}, which showed that for certain nonequilibrium initial states, the system may remain in an instantaneous eigenstate of some local Hamiltonian for times that increase linearly with the particle number~\cite{vidmar_iyer_17, vidmar_xu_17, zhang_vidmar_19, zhang_vidmar_21}.\\

A convenient nonequilibrium setup for observing quasicondensation of HCBs consists of initial domain-wall product states: 
these states refer to lattice configurations where significant parts of the system exhibit fully-filled or empty sites and are connected in a specific manner. 
In particular, a state with HCBs in the center and two domain-wall interfaces can be realized, e.g., by coupling the system to a box potential. 
Since 1D HCBs are integrable, the information about the initial state persists also at times much larger than those in which quasicondensation emerges.
In the expansion of 1D HCBs from a box trap, the initial state information is reflected in the momentum distribution function, which  asymptotically converges to that of spinless fermions, to which the HCBs are mapped~\cite{rigol2005fermionization}. 
This phenomenon is known as dynamical fermionization and is also observed for other integrable systems~\cite{sutherland_98, minguzzi05, bolech12, campbell15, mei16}.

Another important aspect of HCBs lies in their experimental feasibility, both in continuous potentials~\cite{kinoshita04} and in lattice setups~\cite{paredes04, ronzheimer13}. 
These implementations paved way towards the experimental detection of  quasicondensation of 1D HCBs~\cite{vidmar_ronzheimer_15}, as well as to the recent observation of dynamical fermionization~\cite{wilson2020observation}. 
From the theory side, it is hence desirable to develop tools for a complete analytical understanding of far-from-equilibrium dynamics of 1D HCBs. \\

At equilibrium, analytical insights of quasicondensation were obtained by inspecting the behavior of the one-particle density matrix (OPDM), related to the momentum distribution via Fourier transform. Specifically, for homogeneous systems, Luttinger-liquid theory predicts off-diagonal long-range order of the OPDM characterized by power-law decay of spatial correlations. This can be related to quasicondensation considering the scaling of the eigenvalues of the OPDM (Penrose-Onsager criterion \cite{Penrose-Onsager56}) or by investigating the peaks in the momentum distribution e.g.~\cite{rigol2004universal,pezer2007momentum}. Such studies were first extended to the ground state of 1D HCBs in harmonic traps \cite{forrester2003finite}, and more recently to generic inhomogeneous ground-state configurations~\cite{colcelli2018universal, De_Nardis_2018} using an inhomogeneous Luttinger liquid approach~\cite{brun2017one,brun2018inhomogeneous,scopa2020one}. 
Only recently has it become feasible to derive fully analytical expressions for the equal-time one-particle density matrix (OPDM) in generic potentials when the system is out of equilibrium\cite{Scopa_2023}, thanks to a novel quantum hydrodynamic approach. 
The main idea behind the latter is simple: large-scale dynamical correlations can be obtained by building up a non-homogeneous Luttinger liquid field theory on top of the classical hydrodynamic evolution for zero-temperature HCBs~\cite{ruggiero2021quantum,ruggiero2020quantum,scopa2021exact,ruggiero2019conformal}. 
Several analytical results, particularly concerning entanglement dynamics, corroborated the validity of this approach in domain-wall settings, see e.g. \cite{Collura2020domwallmelting,scopa2021exact,ares2022entanglement,scopa2022exact,Rottoli_2022,Scopa2023domainwall,Capizzi_2023}. 
In this work, the focus is on the equal-time OPDM of HCBs released from a box potential. Below, we derive a fully analytical result for the equal-time OBDM, 
allowing us to revisit the emergence of quasicondensation and the development of off-diagonal long-range order in far-from-equilibrium 1D HCBs. \\

\label{outline}\textit{Outline.~---~} The paper is organized as follows. In Sec.~\ref{sec:setup}, we discuss the setup. Namely, we recall the HCBs Hamiltonian and its mapping into spinless free fermions; we specify the initial domain-wall state and the quench protocol that will be investigated. After this, in Sec.~\ref{sec:main-res}, we present our main result for the equal-time OPDM of HCBs during the free expansion. Sec.~\ref{sec:derivation} contains a detailed derivation of our main result. 
This includes the revision of the classical hydrodynamics of HCBs (Sec.~\ref{sec:profile}), its requantization via bosonization (Sec.~\ref{sec:qghd}), and several non-trivial details about the derivation of the OPDM (Sec.~\ref{sec:bosonization}). In Sec.~\ref{sec:applications} we discuss some applications of our results. In particular, we show the the emergence of quasicondensation and of dynamical fermionization at different stages of the dynamics,  exploiting Penrose-Onsager criterion on the largest OPDM eigenvalue and by considering the evolution of the momentum distribution. Finally, Sec.~\ref{sec:conclusion} contains our conclusions. In \ref{app:conjecture} we discuss some asymptotic expansions of form factors that are used in the calculation of the OPDM.

\section{Summary of the results}\label{sec:summary}

\subsection{Setup}\label{sec:setup}
We are interested in the free expansion of a lattice gas of 1D bosons with nearest-neighbor hopping and repulsive on-site interaction $U$, as described by the Bose-Hubbard Hamiltonian
\be\label{eq:BoseHubbard}
\hat{H}=-\frac{1}{2}\sum_x (\hat{a}^\dagger_{x+1}\hat{a}_x +\text{h.c.}) +\frac{U}{2}\sum_x \hat{n}_x(\hat{n}_x-1).
\ee
Here $\hat{a}^\dagger_x$, $\hat{a}_x$ are standard bosonic operators and $\hat{n}_x=\hat{a}^\dagger_x\hat{a}_x$ is the density operator acting on site $x$. Our focus from hereafter is on the limit of strong repulsion $U\gg 1$, where the Hamiltonian \eqref{eq:BoseHubbard} can be written as
\be\label{eq:HCB-Hamiltonian}
\hat{H}=-\frac{1}{2}\sum_x (\hat{b}^\dagger_{x+1}\hat{b}_x +\text{h.c.}) 
\ee
with hard-core bosonic operators $\hat{b}^\dagger_x$, $\hat{b}_x$ satisfying on-site anticommutation relations $\{\hat{b}_x,\hat{b}_{x}^\dagger\}=1$, i.e. avoiding a double occupancy of the same site, while commuting at distinct sites. Identifying each empty/filled lattice site with a down/up spin-$1/2$, the Hamiltonian of HCBs \eqref{eq:HCB-Hamiltonian} is equivalent to the one of the XX spin chain,
\be
\hat{H}  = - \frac{1}{2} \sum_{x \in \mathbb{Z} + \frac{1}{2}}  ( \hat{\sigma}^+_{x} \hat{\sigma}^-_{x+1} + \hat{\sigma}^+_{x+1} \hat{\sigma}^-_{x}  )   = -\frac{1}{4} \sum_{x \in \mathbb{Z} + \frac{1}{2} } \left( \hat{\sigma}_x^x\hat{\sigma}_{x+1}^x + \hat{\sigma}_x^y\hat{\sigma}_{x+1}^y \right),
\label{eq:XX}
\ee
where $\hat{\sigma}_x^a$ is the Pauli operator acting on site $x$ and $\hat{\sigma}^{\pm}_x = ( \hat{\sigma}^{x}_x \pm i \hat{\sigma}^{y}_x  )/2 $. For convenience, we label the lattice sites by half-integers $x = \dots, - \frac{3}{2}, - \frac{1}{2}, \frac{1}{2}, \frac{3}{2}, \dots$. This convention was also used for instance in Ref.~\cite{allegra2016inhomogeneous} and it leads to slightly nicer formulas, as we will see below.

It is well-known that, under the Jordan-Wigner mapping
\be    
    \hat{\sigma}_x^+ =  e^{i \pi \sum_{j = -\infty}^{x-1} \hat{c}_j^\dagger \hat{c}_j} \, \hat{c}_x^\dagger , \qquad
    \hat{\sigma}_x^- = e^{-i \pi \sum_{j = -\infty}^{x-1} \hat{c}_j^\dagger \hat{c}_j}  \,  \hat{c}_x ,
\label{eq:JW_trans}
\ee
where $\hat{c}_x^\dagger$, $\hat{c}_x$, are fermion creation/annihilation operators that obey the canonical anti-commutation relations $\{ \hat{c}_x , \hat{c}_{x'}^\dagger \} = \delta_{x,x'}$, 
the Hamiltonian (\ref{eq:XX}) maps to the one of non-interacting fermions 
\be
\hat{H} = -\frac{1}{2} \sum_{x \in \mathbb{Z} + \frac{1}{2}} \left( \hat{c}^\dagger_x \hat{c}_{x+1} +  \hat{c}^\dagger_{x+1} \hat{c}_{x}  \right). 
\label{eq:hoppingH}
\ee

\begin{figure}[t!]
  \centering
  (a) \hspace{5cm} (b)\\
  \begin{minipage}{.495\textwidth}
 \includegraphics[width=.9\textwidth]{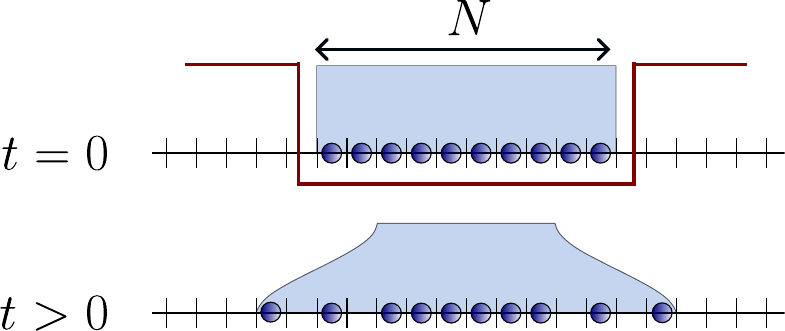}
 \end{minipage}
 \begin{minipage}{.495\textwidth}
 \includegraphics[width=\textwidth]{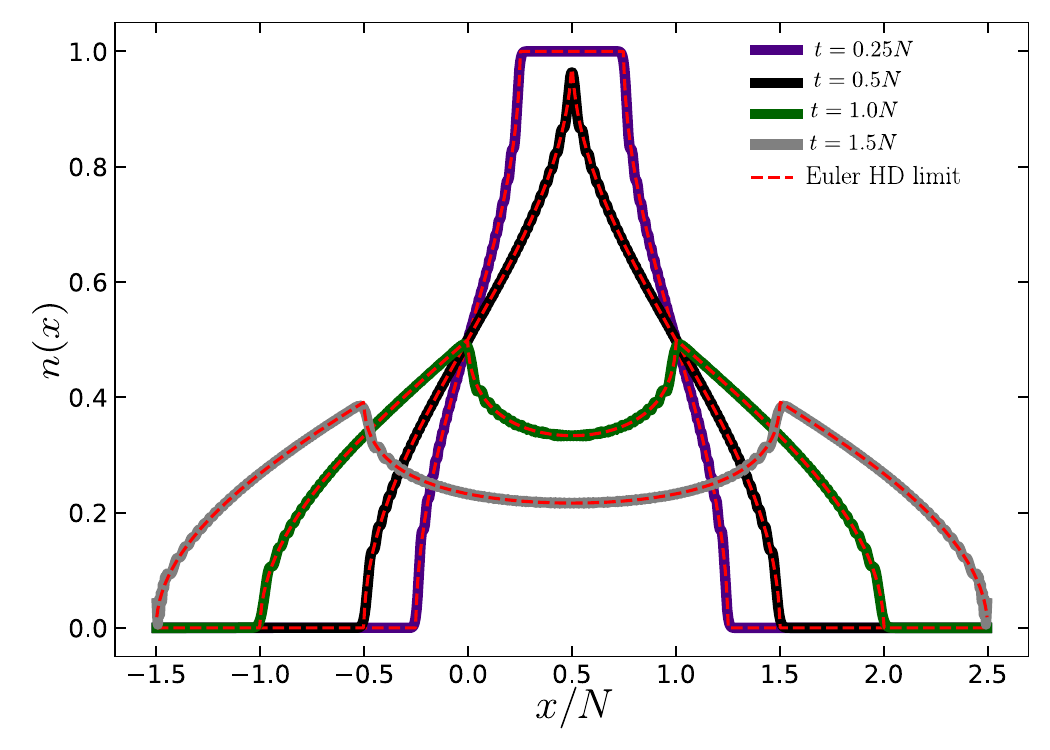}
 \end{minipage}
  \caption{ {(a)}~--~Illustration of the quench protocol considered in this work. At $t=0$, $N$ HCBs are confined by a box potential (eq.~\eqref{eq:potential}) in the region $0<x<N$ of a 1D lattice. At $t>0$, the box potential is switched off and the HCBs freely expands. {(b)}~--~Density profiles $n(x)$ at different times during the box release (see legend). We compare the result of Eq.~\eqref{eq:densityprofile} (dashed lines) with exact numerical calculation (solid lines). The agreement is excellent at all times.}
 \label{fig:lattice}
\end{figure}

In this paper we focus on the following protocol. At time $t = 0$ the system is prepared in the domain-wall product state
\be\label{eq:psi0}
\ket{\psi_0}=\bigotimes_{x<0} \ket{\downarrow_x} \otimes \bigotimes_{0<x<N}\ket{\uparrow_x} \otimes \bigotimes_{x>N} \ket{\downarrow_x}
\ee
where $\ket{\uparrow_x}$ (resp.~$\ket{\downarrow_x})$ are the eigenstates of $\hat\sigma_x^z$ with eigenvalue $+1$ (resp.~$-1$). In the HCBs language, the system contains $N$ particles, with sites $x = \frac{1}{2} , \frac{3}{2} , \dots, N-\frac{1}{2}$ all filled by one hard-core boson while all the others are empty, see Fig.~\ref{fig:lattice}. A possibility to realize the state \eqref{eq:psi0} is to couple the system to a high box potential, namely to consider the initial Hamiltonian
\be
\hat{H}_0=\hat{H}+ \sum_{x \in \mathbb{Z} + \frac{1}{2}} V(x) \hat{c}^\dagger_x\hat{c}_x
\ee
with
\be
V(x) = \begin{cases}
    -V_0  \quad \text{ if } \quad 0 < x < N \\
    +V_0  \qquad \text{ otherwise} ;
\end{cases}
\label{eq:potential}
\ee
and $V_0 \gg 1$, so that $\ket{\psi_0}$ corresponds to the many-particle ground state of $\hat{H_0}$.
This setup has been considered in e.g. Refs.~\cite{Karevski2002,olshanii2003short, rigol2004universal, rigol2007relaxation, Eisler_2021,scopa2022exact}. It has been also experimentally realized with ultracold atoms in optical lattices in Refs. \cite{fang2016momentum, schemmer2019generalized, bouchoule2022generalized}.\\

At time $t>0$, the high box trap is switched off $V(x)=0$, and the HCBs are let to expand freely according to the Hamiltonian (\ref{eq:XX}),
\begin{equation}
    \ket{\psi_t} = e^{- i t\hat{H}} \ket{\psi_0} .
\end{equation}
Using the fact that the dynamics maps to the one of non-interacting fermions under the Jordan-Wigner mapping, the density profile can be easily computed in the Euler hydrodynamic limit \cite{antal1999transport,antal2008log,Karevski2002,Platini_2007,Scopa2023domainwall}. This limit corresponds to
\begin{equation}
    \label{eq:Euler}
    x,t, N \rightarrow \infty  \qquad {\rm with \; ratios} \quad x/t \quad {\rm and} \quad N/t \quad {\rm fixed} .
\end{equation}
The result ---derived e.g. in Refs.~\cite{scopa2022exact}, see also Section~\ref{sec:profile}--- reads
\begin{multline}    \label{eq:densityprofile}
    n(x,t) = \left< \psi_t \right| \frac{1+\hat{\sigma}_x^z}{2} \left| \psi_t \right> = \\
    \left\{ \begin{array}{ccl}
        0 &{\rm if}& x < -t \\
        \frac{1}{\pi} {\rm arccos} (\frac{-x}{t})  &{\rm if}&  -t< x < {\rm min} ( N-t,t) \\
        1 & {\rm if} &  t < \frac{N}{2} \quad {\rm and}  \quad t < x < N-t \\
        \frac{1}{\pi} {\rm arccos}(\frac{x-N}{t}) - \frac{1}{\pi} {\rm arccos} (\frac{x}{t}) & {\rm if} & t > \frac{N}{2} \quad {\rm and}  \quad  N-t < x< t \\  
        \frac{1}{\pi} {\rm arccos} (\frac{x-N}{t})  &{\rm if}&  {\rm max} (N-t, t) < x < N+t \\
        0 &{\rm if}& x > N+t  .
    \end{array} \right.
\end{multline}
The density profile $n(x,t)$ computed exactly for finite $N$ is shown in Fig.~\ref{fig:lattice}; it matches the asymptotic result (\ref{eq:densityprofile}) perfectly.

\subsection{Main result: exact asymptotic formula for the one-particle density matrix}\label{sec:main-res}
The main goal of this paper is to derive an asymptotic formula valid in the Euler limit (i.e., analogous to Eq.~\eqref{eq:densityprofile}) for the equal-time OPDM
\be 
\label{eq:OPDM}
g_1(x,y;t) \, := \, \bra{\psi_t } \hat{\sigma}^+_x \hat{\sigma}^-_y \ket{\psi_t },
\ee
whose Fourier transform is the momentum distribution measured in cold atom experiments, see Refs.~\cite{gerbier2003momentum, regal2005momentum, jukic2009momentum}. 
We will discuss the momentum distribution of the bosons in detail in Sec.~\ref{sec:applications}, but for now let us focus on the quantity (\ref{eq:OPDM}). 
The calculation of the OPDM for HCBs is notoriously hard. The reason is due to its non-local structure in terms of the fermion creation/annihilation operators. Indeed, using the Jordan-Wigner mapping, one finds
\be 
\label{eq:OPDMfermions}
g_1(x,y;t) = \bra{\psi_t } \hat{c}^\dagger_x (-1)^{\sum_j \hat{c}_j^\dagger \hat{c}_j} \hat{c}_y \ket{\psi_t },
\ee
where the sum runs over all sites $j$ between $x$ and $y$. Because the state $\ket{\psi_t}$ is a state of non-interacting fermions ---it can be written as a Slater determinant---, it is possible to use Wick's theorem to evaluate (\ref{eq:OPDMfermions}). However, the Jordan-Wigner string $(-1)^{\sum_j \hat{c}_j^\dagger \hat{c}_j}$ cannot be simplified, so this only results in expressions for $g_1(x,y;t)$ in the form of a determinant of a matrix of size of order $O(N)$. Such large determinants can sometimes be analyzed asymptotically, typically through Toeplitz determinant methods~\cite{vaidya1979one,lenard1964momentum,lenard1972some}. However, this analytical treatment is typically restricted to translation invariant systems. In most situations, in particular in the strongly out-of-equilibrium situation of a box release in which we are interested here, no analytical methods are known and such large determinants need to be evaluated numerically. There exist many works in the literature that are based on such numerical evaluations of $g_1(x,y;t)$ for HCB gases, see for instance Refs.~\cite{rigol2004universal, rigol2007relaxation}.\\

In contrast, in this paper we report a closed expression for $g_1(x,y;t)$ in the Euler limit (\ref{eq:Euler}), expressed entirely in terms of simple functions. Although the formula below may seem somewhat intricate, we stress that it is exceedingly simpler than previously existing expressions \cite{Rigol2004quasicond, rigol2004universal} involving the determinant of a matrix of size $O(N)$.\\

\begin{figure}[t!]
    \centering
    \includegraphics[width = .9\textwidth]{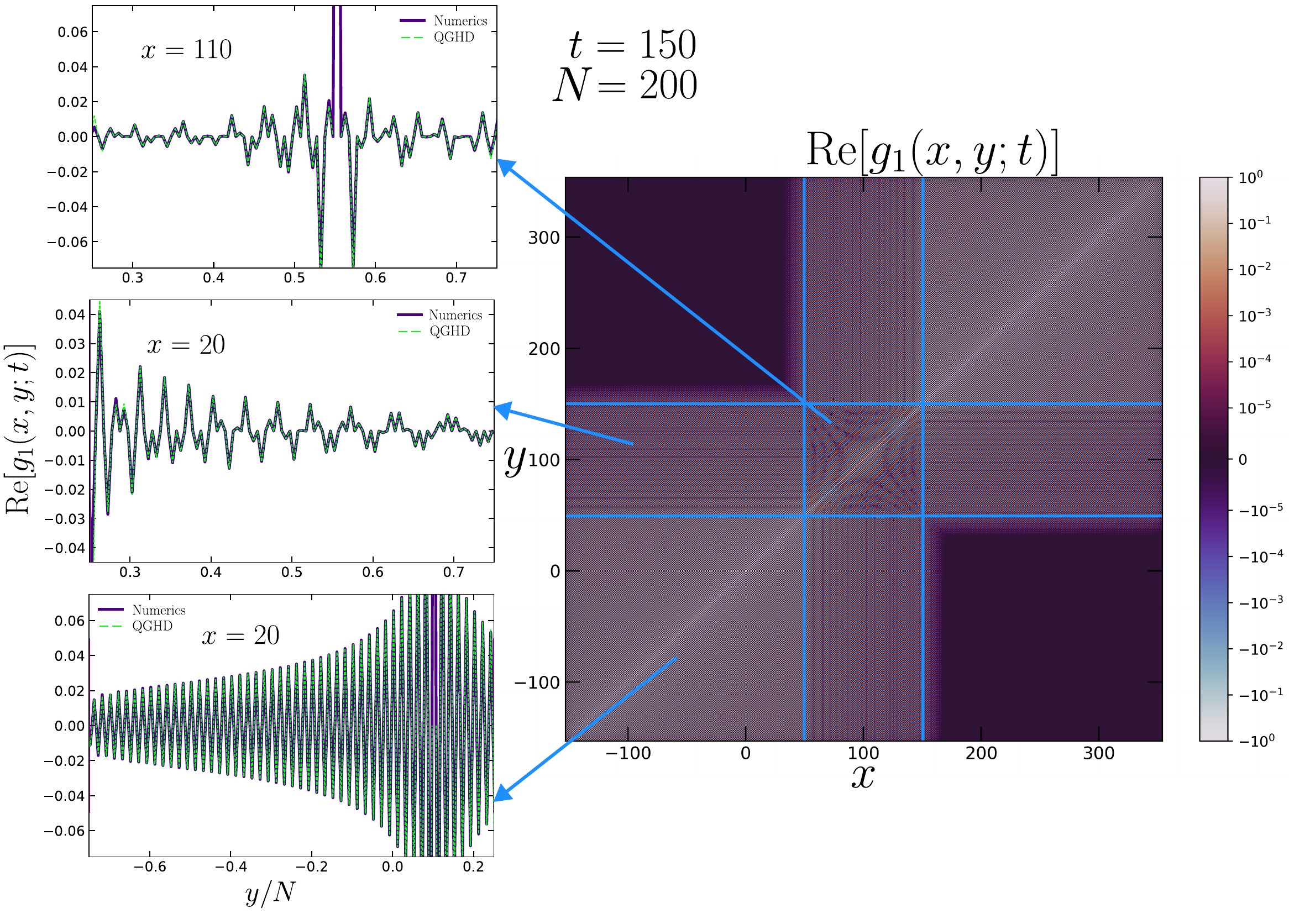}
    \caption{(Right)~--~Color map of $g_1(x,y;t)$ at fixed $t = 150$ and $N = 200$.  Light-blue lines separate the $x$-$y$ plane into 9 different regions, three of which display a qualitatively different behavior, and thus are shown on the left for a fixed value of $x$. (Left)~--~From top to bottom: i) region where positions $x$ and $y$ are both characterized by split Fermi seas~(see discussion in Sec.~\ref{sec:derivation}); ii) region where $x$ correspond to a split Fermi sea while $y$ corresponds to a connected one; iii) region where both $x$ and $y$ correspond to a connected Fermi sea.  Analytical results (dashed lines) and exact numerical calculations (solid lines) are found in perfect agreement.  }
    \label{fig:OPDM_result}
\end{figure}

\begin{figure}[t!]
    \centering
    \includegraphics[width = 0.9\textwidth]{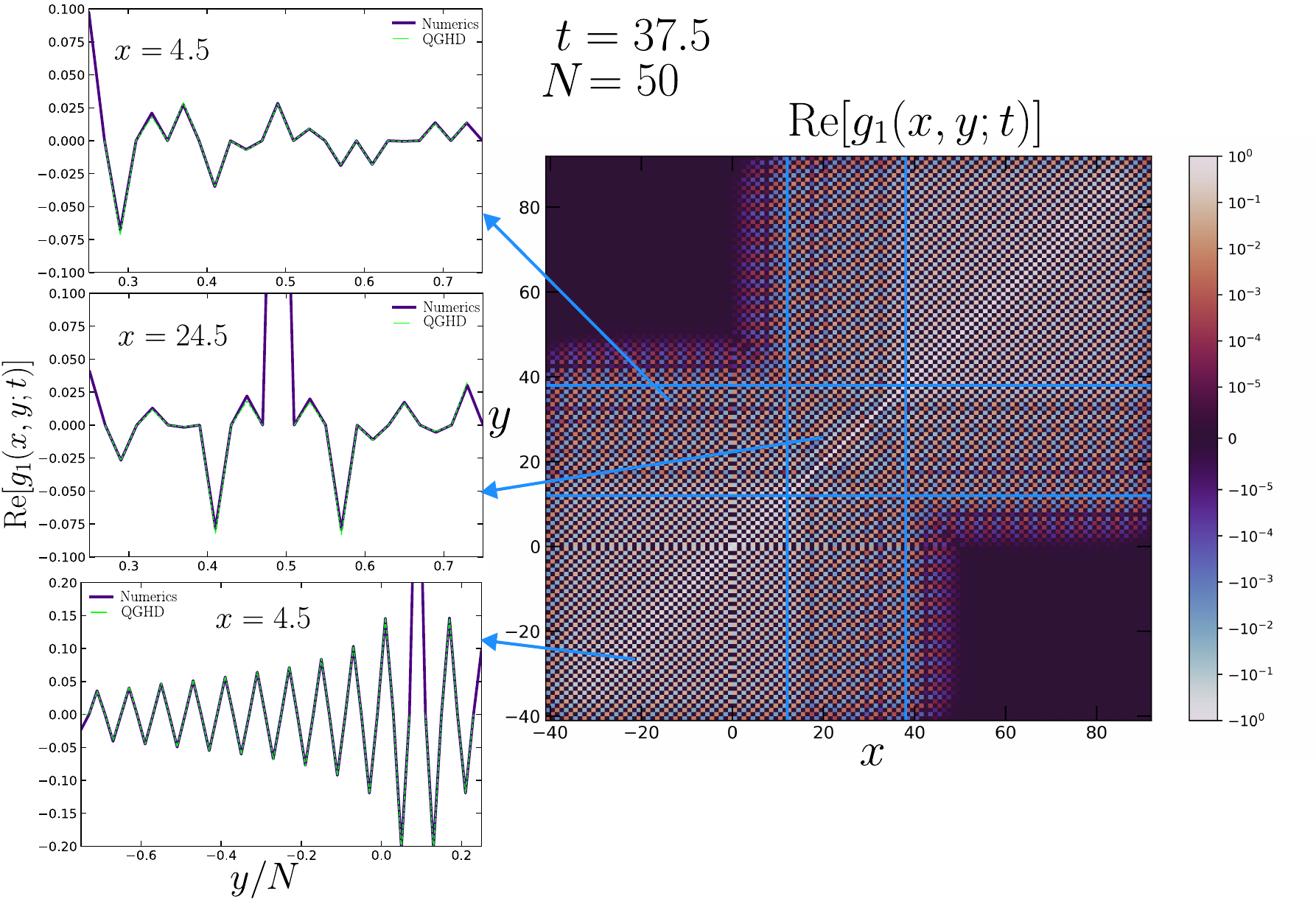}
    \caption{(Right)~--~Another color map of $g_1(x,y;t)$ at $t = 37.5$  for a smaller number of HCBs in the initial state, $N = 50$. Time is set such that the ratio $\frac{N}{t}$ is the same of Fig.~\ref{fig:OPDM_result}. (Left)~--~From top to bottom: i) region where positions $x$ and $y$ are both characterized by split Fermi seas~(see discussion in Sec.~\ref{sec:derivation}) ii) region where $x$ correspond to a split Fermi sea while $y$ corresponds to a connected one; iii) region where both $x$ and $y$ correspond to a connected Fermi sea. Analytical results (dashed lines) and exact numerical calculations (solid lines) are found in perfect agreement. }
    \label{fig:OPDM_result2}
\end{figure}

Now let us present our main result, for which we provide a detailed derivation in Section~\ref{sec:derivation}. It is valid in the  Euler limit $x,y,N,t \rightarrow \infty$, keeping the ratios $\frac{x}{t}$, $\frac{y}{t}$, and $\frac{N}{t}$ fixed. Several cases need to be distinguished, as in formula (\ref{eq:densityprofile}), and this is best summarized in a table. For a given time $t$, we give the following names to the different spatial regions:
\begin{equation}
\begin{tabular}{c|c}
     Interval & Name  \\ \hline \hline
     $x< -t$ or $x>N+t$  &  Empty (E)   \\ &   \\ 
     $-t < x < {\rm min}(t,N-t)$ &  Left (L)  \\ & \\
     $ N-t <x<t $ \quad (if $t>\frac{N}{2}$) &  Central (C)   \\ & \\
    $ {\rm max} (t,N-t) < x < N+t$ &  Right (R)  \\ & \\ 
     $ t<x< N-t$ \quad (if $t< \frac{N}{2}$) &  Filled (F) .
\end{tabular}
\label{eq:regions0}
\end{equation}
Then $g_1(x,y;t)$ is given as follows:
{
\begin{equation}
    \begin{tabular}{c|c|c|c|c|c}
        $g_1(x,y;t)$ & $x\in E$ & $x \in L$ &  $x \in C$ & $x \in R$ & $x\in F$\\  \hline
     
     $y\in E$ & $0$ & $0$ & $0$ & $0$ & $0$\\
     $y\in L$ &  $0$ &  $f(x,y) $ & 
    $g(x,y)$
     & $ 0  $  & $0$\\ 
     $y \in C$ &  $0$ &$(g(y,x))^*$
     &   $h(x,y)$   & 
     $ (g(N-y,N-x))^* $ & $0$  \\
     $y \in R$ &  $0$ & $0$  &
          $g(N-x,N-y)$
     & $f (N-x,N-y)$ & $0$\\
          $y\in F$ & $0$ & $0$ & $0$ & $0$ & $0$\\
    \end{tabular}
    \label{eq:result2}
\end{equation}

We remark that eq.~\eqref{eq:result2} applies to any time (with region $C$ enabled for $t>N/2$), as specified in eq.~\eqref{eq:regions0}}. Here $*$ stands for complex conjugation, and the functions $f$, $g$ and $h$ are defined as 
\begin{eqnarray}
 \nonumber f(x,y) & := &  \frac{G(3/2)^4}{\pi}   e^{i \frac{\pi}{2}(x-y)}  \frac{r(x)^{1/4} r(y)^{1/4}}{(2t)^{1/2}|x-y|^{1/2}}  \\
 \nonumber    g(x,y) & := &
    \frac{G(3/2)^6}{\pi^{3/2}}  \cos\left[ \varphi(N-x)\right] e^{i \frac{\pi}{2}(x-y)} \frac{r(x)^{1/4}r(y)^{1/4}}{r(N-x)^{3/4} |x-y|^{1/2}} , \\
  \nonumber           h(x, y) & := & 
            \frac{G(3/2)^8}{\pi^2}  \left( \cos[\varphi(y) - \varphi(x)] D^{-1/2}(x,y) + \cos[\varphi(y) + \varphi(x)] D^{1/2}(x,y) \right)  \\
 \nonumber            && \times e^{- i\frac{\pi}{2}(x-y)} \frac{(t/2)^{1/2} r(N-x)^{1/4}r(N-y)^{1/4}}{r(x)^{3/4} r(y)^{3/4}|x-y|^{1/2}} \\
            && \qquad + \quad \{ x \rightarrow N-x {\rm , } \quad y \rightarrow N - y \} ,
            \label{eq:h_def}
\end{eqnarray}
where $G(.)$ is the Barnes' $G$ function with $G(3/2) \simeq 1.06922$. Here $\{ x \rightarrow N-x, y \rightarrow N-y \}$ stands for the same term as in the two previous lines with $x$ replaced by $N-x$ and $y$ by $N-y$. We have also defined
\begin{equation}
    \begin{split}
    r(x) & := \, \sqrt{ t^2 - x^2 },  \qquad \varphi(x) \, := \, \sqrt{t^2-x^2} - x\, \arccos (\frac{x}{t}), \\
    D(x,y) & := \, \frac{|x-y|}{t} \frac{1}{1 - \frac{xy}{t^2} + \sqrt{1 - \frac{x^2}{t^2}}\sqrt{1 - \frac{y^2}{t^2}}} \, .
     \end{split}
     \label{eq:supp_funcs}
\end{equation}
We have dropped the explicit dependence on $t$ to lighten the notations. Moreover,
notice that all expressions inside the roots are always positive; we always take the real positive root.\\

In Fig.~\ref{fig:OPDM_result} we compare our asymptotic result to direct numerical calculations of the OPDM (\ref{eq:OPDM}), obtained by evaluating determinants of order $O(N)$~\cite{Rigol2004quasicond, rigol2004universal}. We find that the agreement is excellent. Notice that, despite our formula being derived in the limit $N\to\infty$, it captures with excellent accuracy the behavior of the OPDM already when $N$ is of the order of a few tens, see Fig.~\ref{fig:OPDM_result2}.

\section{One-particle density matrix from `quantum GHD' approach}
\label{sec:derivation}
In this section we explain how we obtain our exact asymptotic formula \eqref{eq:result2} for the equal-time OPDM. We first present the `Generalized Hydrodynamics' (GHD) \cite{bertini2016transport,castro2016emergent} description of HCBs released from the infinite box trap, after Ref.~\cite{scopa2022exact} (see also e.g. Refs.~\cite{Collura_2018,PhysRevA.89.063603, Rottoli_2022,Capizzi_2023b,Collura2020domwallmelting,PhysRevB.99.174403,scopa2021exact,ares2022entanglement,Scopa2023domainwall} for previous works involving an initial state with a single domain wall). We then incorporate quantum fluctuations in the initial state, which we transport with the hydrodynamic equation to later times, following the prescriptions of `inhomogeneous conformal field theory' of Refs.~\cite{dubail2017conformal,allegra2016inhomogeneous,brun2017one,ruggiero2019conformal} or `quantum GHD'~\cite{ruggiero2020quantum,ruggiero2021quantum}.

We stress that our calculation is entirely based on the idea of propagating the quantum fluctuations present in the initial state to later times $t>0$, along with the hydrodynamic (GHD) background. We do not take into account the corrections to the propagation equation coming from the expansion of the Moyal product discussed by Fagotti in Refs.~\cite{fagotti2017higher,fagotti2020locally,Bocini2023}. These corrections should be subleading, and it is not yet understood how to include them in the calculation of the OPDM. 

\subsection{Generalized Hydrodynamics of non-interacting fermions}
\label{sec:profile}

The Hamiltonian (\ref{eq:hoppingH}) is diagonalised by the Fourier transformation
\be
\begin{split}
\hat{c}(k) = \sum_{x \in \mathbb{Z} + \frac{1}{2}} e^{-ikx} \hat{c}_x , \qquad \quad
\hat{c}^\dagger(k) = \sum_{x \in \mathbb{Z} + \frac{1}{2}} e^{ikx} \hat{c}_x^\dagger , 
\label{eq:fourier_modes}
\end{split}
\ee
where $\hat{c}(k)$ ($\hat{c}^\dagger(k)$) annihilates (creates) a particle of momentum $k$, with $k \in \mathbb{R}/2\pi \mathbb{Z}$ (i.e. $k$ is defined modulo $2\pi$):
\be
\hat{H} =  \sum_k \varepsilon(k) \hat{c}^\dagger(k) \hat{c}(k) ,
\ee
with the dispersion relation $\varepsilon(k) = - \cos (k)$.
We do not use the Hamiltonian directly, but rather turn to a hydrodynamic description valid at large distances and large times. In that description the variable $x$ becomes a continuous variable. The state of the gas at time $t$ is described by the phase-space occupation of fermions,
\begin{equation}
    \label{eq:dist_nu}
    \nu(x,k,t) \in [0,1],
\end{equation}
where $(x,k) \in \mathbb{R} \times (\mathbb{R}/2\pi \mathbb{Z})$ are the coordinates in phase-space. It evolves in time according to
\begin{equation}
    \label{eq:ghd}
    \partial_t \nu(x,k,t) + \sin (k) \partial_x \nu(x,k,t)  = 0.
\end{equation}
Here $\sin(k)$ is the group velocity corresponding to the dispersion relation $-\cos(k)$ in the Hamiltonian (\ref{eq:hoppingH}).
The function $\nu(x,k)$ can be thought of as the Wigner function of the fermions, that becomes a classical probability distribution (i.e. with no negative values) in the limit of smooth density variations, see for instance the discussions in Refs.~\cite{doyon2017large,ruggiero2019conformal}. The definition of the Wigner function for lattice systems can be tricky ~\cite{hinarejos2012wigner,fagotti2017higher,fagotti2020locally}; here we simply assume, as in many previous works e.g.~\cite{wendenbaum2013hydro,antal1999transport,Scopa2023domainwall}, that the gas of hard-core bosons at large scales is described by the large-scale occupation (\ref{eq:dist_nu}), which obeys the evolution equation (\ref{eq:ghd}).


The initial occupation corresponding to our initial state (\ref{eq:psi0}) is
\be
\nu (x,k,0) = \begin{cases}
    1 \quad \text{ if } \quad 0 < x < N , \\
    0 \quad \text{ otherwise}.
\end{cases}
\ee
This situation, where the initial occupation $\nu(x,k,0)$ is either $0$ or $1$, is a special case known as `Generalized Hydrodynamics for zero-entropy states' or `zero-entropy GHD'~\cite{doyon2017large}. In that case, it is easy to see that the solution of the evolution equation (\ref{eq:ghd}) is
\begin{equation}
    \nu(x,k,t) = \left\{ 
        \begin{array}{rcc}
            1 & {\rm if} &  t \sin k < x < N + t \sin k \\
            0 & & {\rm otherwise} . 
        \end{array}
    \right.
\end{equation}
The particle density at position $x$ is obtained by integrating the phase-space occupation over $k$,
\begin{equation}
    \label{eq:density}
    n(x,t) = \int_{-\pi}^\pi \nu(x,k,t) \frac{dk}{2\pi} ,
\end{equation}
which reproduces the anticipated formula (\ref{eq:densityprofile}) for the density profile.

\begin{figure}
    \centering
    \includegraphics[width = .55\textwidth]{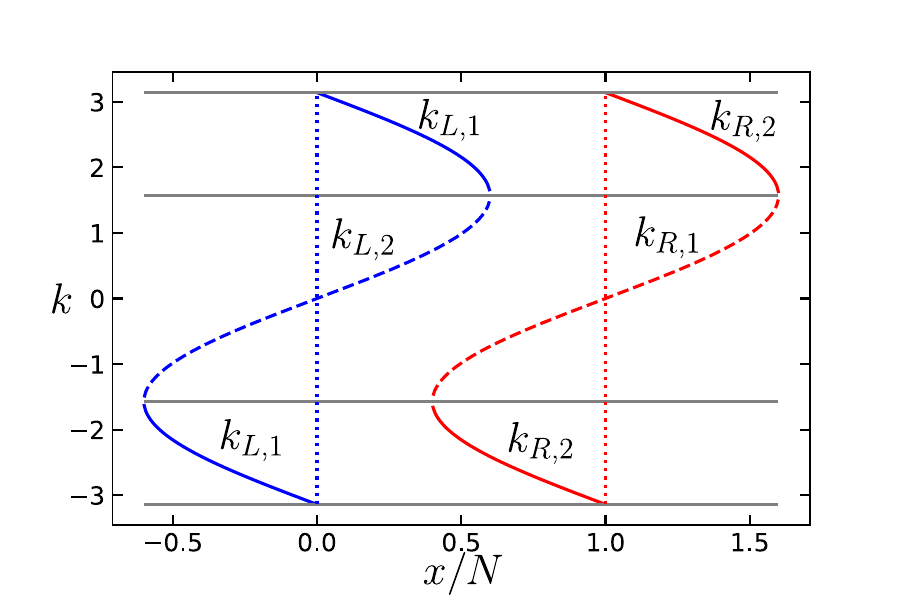}
    \caption{Parametrization of the left/right Fermi contours for a system with $N = 200$ HCBs at $t = 120$. The blue (resp.~red) solid line shows $k_{L,1}(x)$ (resp.~$k_{R,2}(x)$), while the blue (resp.~red) dashed line shows $k_{L,2}(x)$ (resp.~$k_{R,1}(x)$). Dotted lines show the Fermi contours at $t = 0$ (blue~--~left; red~--~right); horizontal grey lines indicate the momenta $k=\pi$, $\frac{\pi}{2}$,  $- \frac{\pi}{2}$ and $-\pi$ from top to bottom respectively.}
    \label{fig:Fermi_cont_param}
\end{figure}

The two curves that separate the phase-space region where $\nu=1$ from the ones where $\nu = 0$ play a central role in this work. We refer to them as the left and right `Fermi contours'. At a fixed time $t$, they are defined as
\begin{eqnarray}
    \label{eq:contour_s}
    \nonumber \Gamma_{\rm L} &:=& \left\{ (x,k) = (N \sin s, s) , \; s \in [-\pi, \pi] \right\} ,  \\  \Gamma_{\rm R} &:=& \left\{ (x,k) = (N+t \sin s, s) , \; s \in [-\pi, \pi] \right\} .
\end{eqnarray}
Below we will sometimes use the following parametrization (see Fig.~\ref{fig:Fermi_cont_param}):
 \begin{eqnarray}
    \label{eq:parameterization}
\nonumber k_{{\rm L},1}(x) \, := \, \left( \pi - \arcsin \left( \frac{|x|}{t}\right)\right) \text{sign}(x) & \in & [-\pi, -\frac{\pi}{2}] \cup [\frac{\pi}{2}, \pi]   \\
\nonumber k_{{\rm L},2}(x) \, := \,  \arcsin \left( \frac{x}{t}\right) &\in& [- \frac{\pi}{2}, \frac{\pi}{2}] \\
\nonumber  k_{{\rm R}, 1}(x) \, :=  \,   \arcsin\left( \frac{x-N}{t} \right) & \in & [-\frac{\pi}{2}, \frac{\pi}{2}] \\
\nonumber  k_{{\rm R}, 2}(x) \, := \,  \left( \pi - \arcsin\left( \frac{|x-N|}{t} \right) \right) \sign( x - N )  & \in &  [-\pi, -\frac{\pi}{2}] \cup [\frac{\pi}{2}, \pi] , \\
\end{eqnarray}
so that
\begin{eqnarray}
\fl \nonumber    \Gamma_{\rm L} = \left\{ (x,k_{{\rm L},1}(x) ) , \;  x \in [-t,t] \right\} \cup \left\{ (x,k_{{\rm L},2}(x) ) , \;  x \in [-t,t] \right\}  \\  \fl \Gamma_{\rm R} = \left\{ (x,k_{{\rm R},1}(x) ) , \;  x \in [N-t,N+t] \right\} \cup \left\{ (x,k_{{\rm R},2}(x) ) , \;  x \in [N-t,N+t] \right\}  . \qquad \quad 
\end{eqnarray}

We stress that the description of the dynamics of the gas in terms of the occupation ratio (\ref{eq:dist_nu}) and the evolution equation (\ref{eq:ghd}) introduced so far is entirely classical. It does capture some quantities, for instance the density profile (\ref{eq:densityprofile}), in the limit of large time and large distances. However, it does not give access to other quantities like the OPDM (\ref{eq:OPDM}). We now turn to the approach of Refs.~\cite{dubail2017conformal,ruggiero2019conformal,ruggiero2020quantum,Collura2020domwallmelting,ruggiero2021quantum,scopa2021exact,scopa2022exact,Scopa_2023} and reintroduce some `quantumness' into the description, which will give us access to the OPDM.

\subsection{Field theory description of quantum fluctuations}
\label{sec:qghd}
In our description of the expansion of the HCB gas, the `quantumness' of the system resides in the initial state, and is then propagated in time according to the classical equation (\ref{eq:ghd}). Here we explain how this approach is implemented, and how it leads to our main result, which is our asymptotic formula for the OPDM.

We proceed as follows. We start by analyzing the fermionic two-point function at time $t$, which can easily be computed exactly and expressed as a double-integral that can be evaluated by the stationary phase approximation. Then we observe that the result is exactly what we get in a quantum theory describing quantum fluctuations of the two Fermi contours $\Gamma_{\rm L}$ and $\Gamma_{\rm R}$.

\subsubsection{Fermionic two-point function.}

Let us compute the correlation function
\begin{equation}
    \label{eq:fermion2pts}
    \left<\psi_t \right| \hat{c}_x^\dagger \hat{c}_y \left| \psi_t \right> 
\end{equation} 
for two points $x$ and $y$ on the lattice. At time $t=0$, we have (see Eq.~(\ref{eq:psi0}))
\be
\left< \psi_0 \right| \hat{c}_x^\dagger \hat{c}_{y} \left| \psi_0 \right>  = 
\begin{cases}
    \delta_{x,y} \quad \text{ if } \quad 0 < x < N ,  \\
    0 \quad  \text{otherwise}.
\end{cases}
\label{eq:expval_xy}
\ee
In momentum space, this is
\begin{align}
\nonumber 
&\left< \psi_0 \right| \hat{c}^\dagger(k) \hat{c}(q) \left| \psi_0 \right>  =  \sum_{x,y \in \mathbb{Z} + \frac{1}{2}} e^{i(kx-qy)}  \left< \psi_0 \right| \hat{c}^\dagger_x \hat{c}_y \left| \psi_0 \right> \\
\nonumber 
&\quad= \sum_{x = \frac{1}{2}}^{N - \frac{1}{2}}  e^{i(k-q)x} = -2\pi \delta (k-q) + \sum_{x = \frac{1}{2}}^{+\infty}  e^{i(k-q + i 0^+)x} + \sum_{x = -\infty}^{N - \frac{1}{2}}  e^{i(k-q - i 0^+)x}  \\
 &\quad = -2\pi \delta(k-q) +  \frac{-1}{2i \sin \left( \frac{k-q}{2} +i 0^+ \right)} + \frac{e^{i(k-q)N}}{2i \sin \left( \frac{k-q}{2} - i 0^-\right)} .
\label{eq:ckck'}
\end{align}
Then at later time $t$, we use the fact that, for the Hamiltonian (\ref{eq:hoppingH}),
\be
e^{i\hat{H}t} \hat{c}^\dagger(k) e^{-i\hat{H}t} = e^{i\varepsilon(k)t} \hat{c}^\dagger(k) ,
\ee
so that
\be 
\left< \psi_t \right| \hat{c}^\dagger(k) \hat{c}(q) \left| \psi_t \right> = e^{i t [ \varepsilon(k) - \varepsilon(q)]} \left< \psi_0 \right| \hat{c}^\dagger(k) \hat{c}(q) \left| \psi_0 \right>.
\label{eq:ckck'_t}
\ee
This leads to the double-integral
\begin{eqnarray}
\label{eq:Fourier_trans}
\nonumber \left< \psi_t \right| \hat{c}^\dagger_x \hat{c}_y \left| \psi_t \right> &=& \int_{- \pi}^{\pi} \frac{dk}{2\pi} \int_{- \pi}^{\pi} \frac{dq}{2\pi} e^{-i ( kx - qy) } \left< \psi_t \right| \hat{c}^\dagger(k) \hat{c}(q) \left|\psi_t \right> \\
\nonumber &=& -\delta_{x,y} + \int_{- \pi}^{\pi} \frac{dk}{2\pi} \int_{- \pi}^{\pi} \frac{dq}{2\pi} e^{-i ( kx - \varepsilon(k) t ) } e^{i ( qy - \varepsilon(q) t ) } \\
&& \qquad \left( \frac{-1}{2i \sin \left( \frac{k-q}{2} +i 0^+ \right)}  + \frac{e^{i(k-q)N}}{2i \sin \left( \frac{k-q}{2} - i 0^+ \right)} \right) . \qquad
\end{eqnarray}
In the limit of large time $t$, keeping the ratios $x/t$ and $y/t$ fixed, this double-integral can be evaluated using the stationary phase approximation. For the first term in the parenthesis, the phase  is stationary when $k = k_{{\rm L},1}$ or $k = k_{{\rm L},2}$, so that
\be
    \label{eq:stationary_point}
x - \varepsilon'(k)t = 0 .
\ee
Expanding the phase around the stationary point $k_{{\rm L}, a}(x)$ ($a=1,2$), one gets
\begin{eqnarray}
\nonumber k x - \varepsilon(k)t & \simeq & \Phi(k_{{\rm L}, a}(x)) - \frac{1}{2} \varepsilon''(k_{{\rm L},a}(x)) \, t \, (k - k_{{\rm L},a}(x))^2   \\
 & = &  \Phi(k_{{\rm L}, a}(x)) - \frac{1}{2} \left( \frac{dk_{{\rm L},a}(x) }{dx} \right)^{-1} (k - k_{{\rm L},a}(x))^2 ,
\end{eqnarray}
where we have used that $dk_{{\rm L},a}/dx = 1/ (\varepsilon''(k_{{\rm L},a}) t)$, which follows from Eq.~(\ref{eq:stationary_point}), and we have defined the `WKB phase' (Brioullin-Kramers-Wentzel) as
\begin{equation}
    \label{eq:WKB}
    \Phi(k) \, := \, k   t \sin k - \varepsilon(k)t .
\end{equation}
We have a similar expansion for the second term in the parenthesis in Eq.~(\ref{eq:Fourier_trans}), around the points $k = k_{{\rm R},1}(x)$ and $k = k_{{\rm R},2}(x)$, which solve the stationary phase condition $
x-N - \varepsilon'(k)t = 0$. Around that point the expansion reads
\be
k (x-N) - \varepsilon(k)t \approx \Phi(k_{{\rm R}, a}(x)) - \frac{1}{2} \left( \frac{dk_{{\rm R},a}(x)}{dx} \right)^{-1} (k - k_{{\rm R},a}(x))^2  ,
\ee
with the same function $\Phi(k)$ as in Eq.~(\ref{eq:WKB}). Notice that there may be no stationary points, depending on the region where the point $x$ is, see the table (\ref{eq:regions0}).

Let us first focus on the first term in the integral in Eq.~(\ref{eq:Fourier_trans}), assuming that $-t < x,y < t$ so that both $x$ and $y$ are in the left (L) or central (C) region. Then there are two stationary points for both integrals, over $k$ and over $q$. Evaluating the Gaussian integrals, we arrive at
 \begin{align}
\nonumber &\quad \int_{- \pi}^{\pi} \frac{dk}{2\pi} \int_{- \pi}^{\pi} \frac{dq}{2\pi} e^{-i ( kx - \varepsilon(k) t ) } e^{i ( qy - \varepsilon(q) t ) }  \frac{-1}{2i \sin \left( \frac{k-q}{2} + i 0^+\right)}   \\
\nonumber 
&\simeq  \sum_{a=1,2} \sum_{b=1,2} \int_{- \infty}^{\infty} \frac{dk}{2\pi} \int_{- \infty}^{\infty} \frac{dq}{2\pi} \frac{ e^{-i \Phi ( k_{{\rm L}, a}(x) ) + \frac{i}{2}(k-k_{{\rm L}, a}(x))^2  \varepsilon''(k_{{\rm L}, a}) t } e^{i \Phi ( k_{{\rm L}, b}(y) ) - \frac{i}{2}(k-k_{{\rm L}, b}(y))^2  \varepsilon''(k_{{\rm L}, b}) t } }{ -2i \sin \left( \frac{k_{{\rm L},a}(x)-k_{{\rm L},b}(y) }{2} \right)} \\
&= \sum_{a,b}  \frac{1}{2\pi}   \sqrt{\frac{d k_{{\rm L}, a}(x)}{dx}} \sqrt{\frac{d k_{{\rm L}, b}(y)}{dy}} \frac{i \, e^{- i [ \Phi (k_{{\rm L},a} (x)) - \Phi (k_{{\rm L},b} (y)) ] }}{2 \sin \left( \frac{k_{{\rm L}, a}(x)-k_{{\rm L}, b}(y)}{2} \right)},
\label{eq:corr_final1}
\end{align}
where if the expression inside the square roots is negative we choose the branches in the following way:
\begin{eqnarray}
     \nonumber {\rm Im}\left[ \sqrt{\frac{d k_{{\rm L}, a}(x)}{dx}} \right] < 0, \quad {\rm if } \quad \frac{d k_{{\rm L}, a}(x)}{dx} < 0 \qquad {\rm i.e. \quad if \;} a=1, \qquad {\rm and } \\ 
     {\rm Im}\left[ \sqrt{\frac{d k_{{\rm L}, b}(y)}{dy}} \right] > 0, \quad {\rm if } \quad \frac{d k_{{\rm L}, b}(x)}{dx} < 0 \qquad {\rm i.e. \quad if \;} b=1.
     \label{eq:branches}
\end{eqnarray}
Similarly, for the second term in the parenthesis in Eq.~(\ref{eq:Fourier_trans}), we arrive at
 \begin{align}
\nonumber &\quad \int_{- \pi}^{\pi} \frac{dk}{2\pi} \int_{- \pi}^{\pi} \frac{dq}{2\pi} e^{-i ( kx - \varepsilon(k) t ) } e^{i ( qy - \varepsilon(q) t ) }   \frac{e^{i(k-q)N}}{2i \sin \left( \frac{k-q}{2} - i 0^+ \right)}    \\
\nonumber &\simeq  \sum_{a=1,2} \sum_{b=1,2} \int_{- \infty}^{\infty} \frac{dk}{2\pi} \int_{- \infty}^{\infty} \frac{dq}{2\pi} \frac{ e^{-i \Phi ( k_{{\rm R}, a}(x) ) + \frac{i}{2}(k-k_{{\rm R}, a}(x))^2  \varepsilon''(k_{{\rm R}, a}(x)) t } e^{i \Phi ( k_{{\rm R}, b}(y) ) - \frac{i}{2}(k-k_{{\rm R}, b})^2  \varepsilon''(k_{{\rm R}, b}) t } }{ 2i \sin \left( \frac{k_{{\rm R},a}(x)-k_{{\rm R},b}(y) }{2} \right)} \\
&=\sum_{a,b}  \frac{1}{2\pi}   \sqrt{\frac{d k_{{\rm R}, a}(x)}{dx}} \sqrt{\frac{d k_{{\rm R}, b}(y)}{dy}} \frac{-ie^{- i [ \Phi (k_{{\rm R},a} (x)) - \Phi (k_{{\rm R},b} (y)) ] }}{2 \sin \left( \frac{k_{{\rm R}, a}(x)-k_{{\rm R}, b}(y)}{2} \right)},
\label{eq:corr_final2}
\end{align}
where we choose the branches the same way as we did in Eq.~(\ref{eq:branches}).

\subsubsection{Identification with field theory correlator.}

The key point is to observe that the above expressions for the different terms contributing to $\left< \psi_t \right| \hat{c}^\dagger_x \hat{c}_y \left| \psi_t \right>$ precisely coincide with correlators in a field theory defined along the two Fermi contours. To see this, we introduce fermion fields associated to the L and R Fermi contours,
\begin{equation}
    \label{eq:fermion_fields}
    \psi_{\rm L}^\dagger(k) , \qquad  \psi_{\rm L}(k) , \qquad  \psi_{\rm R}^\dagger(k)  ,  \qquad \psi_{\rm R}(k)  ,
\end{equation}
for which we postulate the following two-point correlation functions,
\begin{equation}
    \label{eq:2point_psi}
    \langle \psi_{\rm L}^\dagger(k)  \psi_{\rm L}(q)  \rangle := \frac{i \, e^{-i [ \Phi_{\rm L}(k) -\Phi_{\rm L}(q)] }}{2 \sin \left( \frac{k-q}{2} +i 0^+ \right) } , \qquad \langle \psi_{\rm R}^\dagger(k) \psi_{\rm R}(q)  \rangle := \frac{-i \, e^{-i [ \Phi_{\rm R}(k) -\Phi_{\rm R}(q)] }}{2 \sin \left( \frac{k-q}{2} - i 0^+\right) }.
\end{equation}
All other two-point functions vanish. Then we can express the fermion creation/annihilation operators in the lattice model in terms of these fields as follows: 
\begin{equation}
    \begin{tabular}{c|c}
         &  $\hat{c}^\dagger_x$  \\ \hline
         $x \in {\rm E}$ &  $0$  \\ \\
         $x \in {\rm L}$ &  $  \displaystyle \frac{e^{-i\frac{\pi}{4}}}{(2\pi)^{\frac{1}{2}}} \left| \frac{dk_{{\rm L},1}}{dx} \right|^{\frac{1}{2}} \psi^\dagger_{{\rm L}}(k_{{\rm L},1}(x)) +  \frac{e^{i\frac{\pi}{4}}}{(2\pi)^{\frac{1}{2}}} \left| \frac{dk_{{\rm L},2}}{dx} \right|^{\frac{1}{2}} \psi^\dagger_{{\rm L}}(k_{{\rm L},2}(x))$ \\ \\
         $x \in {\rm R}$ &  $  \displaystyle \frac{e^{i\frac{\pi}{4}}}{(2\pi)^{\frac{1}{2}}} \left| \frac{dk_{{\rm R},1}}{dx} \right|^{\frac{1}{2}} \psi^\dagger_{{\rm R}}(k_{{\rm R},1}(x)) + \frac{e^{-i\frac{\pi}{4}}}{(2\pi)^{\frac{1}{2}}} \left| \frac{dk_{{\rm R},2}}{dx} \right|^{\frac{1}{2}} \psi^\dagger_{{\rm R}}(k_{{\rm R},2}(x))$ \\ \\
         $x \in {\rm C}$ &  $  \displaystyle \frac{e^{-i\frac{\pi}{4}}}{(2\pi)^{\frac{1}{2}}} \left| \frac{dk_{{\rm L},1}}{dx} \right|^{\frac{1}{2}} \psi^\dagger_{{\rm L}}(k_{{\rm L},1}(x)) + \frac{e^{i\frac{\pi}{4}}}{(2\pi)^{\frac{1}{2}}} \left| \frac{dk_{{\rm L},2}}{dx} \right|^{\frac{1}{2}} \psi^\dagger_{{\rm L}}(k_{{\rm L},2}(x))$ \\
          &  $  \displaystyle +\quad  \frac{e^{i\frac{\pi}{4}}}{(2\pi)^{\frac{1}{2}}} \left| \frac{dk_{{\rm R},1}}{dx} \right|^{\frac{1}{2}} \psi^\dagger_{{\rm R}}(k_{{\rm R},1}(x)) + \frac{e^{-i\frac{\pi}{4}}}{(2\pi)^{\frac{1}{2}}} \left| \frac{dk_{{\rm R},2}}{dx} \right|^{\frac{1}{2}} \psi^\dagger_{{\rm R}}(k_{{\rm R},2}(x))$
    \end{tabular}
\end{equation}
and 
\begin{equation}
    \begin{tabular}{c|c}
         &  $\hat{c}_x$  \\ \hline
         $x \in {\rm E}$ &  $0$  \\ \\
         $x \in {\rm L}$ &  $  \displaystyle \frac{e^{i\frac{\pi}{4}}}{(2\pi)^{\frac{1}{2}}} \left| \frac{dk_{{\rm L},1}}{dx} \right|^{\frac{1}{2}} \psi_{{\rm L}}(k_{{\rm L},1}(x)) + \frac{e^{-i\frac{\pi}{4}}}{(2\pi)^{\frac{1}{2}}} \left| \frac{dk_{{\rm L},2}}{dx} \right|^{\frac{1}{2}} \psi_{{\rm L}}(k_{{\rm L},2}(x))$ \\ \\
         $x \in {\rm R}$ &  $  \displaystyle \frac{e^{-i\frac{\pi}{4}}}{(2\pi)^{\frac{1}{2}}} \left| \frac{dk_{{\rm R},1}}{dx} \right|^{\frac{1}{2}} \psi_{{\rm R}}(k_{{\rm R},1}(x)) + \frac{e^{i\frac{\pi}{4}}}{(2\pi)^{\frac{1}{2}}} \left| \frac{dk_{{\rm R},2}}{dx} \right|^{\frac{1}{2}} \psi_{{\rm R}}(k_{{\rm R},2}(x))$ \\ \\
         $x \in {\rm C}$ &  $  \displaystyle \frac{e^{i\frac{\pi}{4}}}{(2\pi)^{\frac{1}{2}}} \left| \frac{dk_{{\rm L},1}}{dx} \right|^{\frac{1}{2}} \psi_{{\rm L}}(k_{{\rm L},1}(x)) + \frac{e^{-i\frac{\pi}{4}}}{(2\pi)^{\frac{1}{2}}} \left| \frac{dk_{{\rm L},2}}{dx} \right|^{\frac{1}{2}} \psi_{{\rm L}}(k_{{\rm L},2}(x))$ \\
          &  $  \displaystyle +\quad  \frac{e^{-i\frac{\pi}{4}}}{(2\pi)^{\frac{1}{2}}} \left| \frac{dk_{{\rm R},1}}{dx} \right|^{\frac{1}{2}} \psi_{{\rm R}}(k_{{\rm R},1}(x)) + \frac{e^{i\frac{\pi}{4}}}{(2\pi)^{\frac{1}{2 }}} \left| \frac{dk_{{\rm R},2}}{dx} \right|^{\frac{1}{2}} \psi_{{\rm R}}(k_{{\rm R},2}(x))$ .
    \end{tabular}
\end{equation}
These expressions for $\hat{c}^\dagger_x$ and $\hat{c}_x$ are constructed so that, evaluating the correlation function $\left< \psi_t \right| \hat{c}^\dagger_x \hat{c}_y \left| \psi_t \right>$ using Eq.~(\ref{eq:2point_psi}), one automatically reproduces the results obtained above from the stationary phase approximation, see Eqs.~(\ref{eq:corr_final1})-(\ref{eq:corr_final2}).

\subsubsection{Field theory of quantum fluctuations around the Fermi contours.} We have identified the fermionic two-point function at time $t$, $\left< \psi_t \right| \hat{c}^\dagger_x \hat{c}_y \left| \psi_t \right>$, with correlators in a free fermion field theory. Importantly, this identification extends to all higher-point correlation functions of fermionic operators, $\left< \psi_t \right| \hat{c}^\dagger_{x_1} \dots \hat{c}^\dagger_{x_n} \hat{c}_{y_1} \dots \hat{c}_{y_n} \left| \psi_t \right>$, through Wick's theorem. Therefore, all fermionic correlation functions are captured, in the  asymptotic regime $t \rightarrow \infty$ with fixed $\frac{x}{t}$ and $\frac{N}{t}$, by the free fermion field theory defined by the two-point function (\ref{eq:2point_psi}).\\

In the following, we will be interested in the OPDM, which is not easily expressible in terms of a simple fermionic correlation function. Therefore, it is useful to bosonize the field theory. To do this, we introduce two chiral boson fields, defined along the left (L) and right (R) Fermi contours respectively,
\begin{equation}
    \phi_{\rm L}(k), \qquad   \phi_{\rm R}(k),
\end{equation}
and we postulate that these two fields have Gaussian correlation functions, with the following one- and two-point connected functions,
\begin{eqnarray}
 \nonumber   \langle \phi_{\alpha} (k) \rangle &=& \Phi(k) , \qquad \alpha = {\rm L}, {\rm R}, \\
\langle \phi_{{\rm L}}(k) \phi_{{\rm L}}(q) \rangle_{\rm conn.} &=& - \log \left( 2 \sin ( \frac{k - q}{2} + i 0^+ ) \right) ,  \\
   \langle \phi_{{\rm R}}(k) \phi_{{\rm R}}(q) \rangle_{\rm conn.} &=& - \log \left( 2 \sin ( \frac{k - q}{2} - i 0^+ ) \right)  , \qquad  
    \label{eq:phi_correl_LR}
\end{eqnarray}
and $\langle \phi_{{\rm R}}(k) \phi_{{\rm L}}(q) \rangle_{\rm conn.} = 0$. Here $\Phi$ is the WKB phase defined in Eq.~(\ref{eq:WKB}).

In terms of these chiral boson fields, the fermion fields (\ref{eq:fermion_fields}) are given by the standard bosonization formulas
\begin{eqnarray}
    \label{eq:bosonization}
        && \psi_{{\rm L}}^\dagger(k) = \, e^{i \frac{\pi}{4}} :e^{-i \phi_{{\rm L}} (k)}: ,  \qquad \psi_{{\rm L}}(k) = \, e^{i \frac{\pi}{4}} :e^{i \phi_{{\rm L}} (k)}: , \\
    && \psi_{{\rm R}}^\dagger(k) = \, e^{-i \frac{\pi}{4}} :e^{-i \phi_{{\rm R}} (k)}: ,  \qquad \psi_{{\rm R}}(k) = \, e^{-i \frac{\pi}{4}} :e^{i \phi_{{\rm R}} (k)}: , 
\end{eqnarray}
where $:.:$ stands for normal ordering. The chiral bosons $\phi_{\rm R}$ and $\phi_{\rm L}$ can be interpreted as quantum fluctuations of the Fermi contours. Indeed, the following density operators measure the excess of fermions near the L and R contours:
\begin{eqnarray}
    \label{eq:densityfluct1}
    \delta \rho_{\rm L}(k)  \, = \;    : \psi_{\rm L}^\dagger(k) \psi_{\rm L}(k) : \; = \, \partial_k \phi_{\rm L} (k) - \partial_k \langle \phi_{\rm L} (k) \rangle \, , \\
        \delta \rho_{\rm R}(k)  \, = \;    : \psi_{\rm R}^\dagger(k) \psi_{\rm R}(k) : \; = \, -\partial_k \phi_{\rm R} (k) + \partial_k \langle \phi_{\rm R} (k) \rangle \, .
    \label{eq:densityfluct2}
\end{eqnarray}
This is the idea of the theory of `quantum Generalized Hydrodynamics'~\cite{ruggiero2020quantum,ruggiero2021quantum,scopa2021exact,Scopa_2023}: long wavelength quantum fluctuations of the fluid come from small fluctuations of the Fermi contour, encoded as a free chiral boson theory along that contour, with an operator $\delta \rho_\alpha$ that measures the excess density of quasi-particles around a point on the contour. The effective theory describing those fluctuations of the Fermi contour is then similar to a Luttinger liquid.

\subsection{The boson creation/annihilation operator}\label{sec:bosonization}

\subsubsection{Expansion in terms of field theory operators.}
In order to compute the OPDM (\ref{eq:OPDM}),  we need to express the operators that create/annihilate a boson, $\hat{\sigma}^+_x$/$\hat{\sigma}^-_x$, in terms of local operators in the field theory that describe fluctuations of the contour, similarly to what is done in Ref.~\cite{Scopa_2023},
\begin{eqnarray}
    \label{eq:exp_sigma}
\nonumber \hat{\sigma}_x^+  & = & \sum_{\eta \in \mathcal{I}_x } \mathcal{C}^*_\eta(x)  \mathcal{O}^\dagger_\eta(x) , \\
\hat{\sigma}_x^-  & = & \sum_{\eta \in \mathcal{I}_x} \mathcal{C}_\eta(x)   \mathcal{O}_\eta(x) .
\end{eqnarray}
Here the $\mathcal{O}_\eta(x)$'s are operators in the free boson theory, the $\mathcal{C}(x)$'s are complex coefficients, and $\eta$ is a multi-index labeling the operators (see below); $\eta$ runs over a set $\mathcal{I}_x$ that depends on the number of Fermi points at position $x$. In principle, infinitely many field theory operators can appear in the r.h.s. of (\ref{eq:exp_sigma}). However, in practice, these operators can be organized according to their scaling dimension, and the sum can be truncated to keep only the operators of lowest scaling dimension, which are those fixing the leading behavior of  $g_1(x,y;t)$ at large distances. The operators ${\cal O}_\eta$ are primary operators of the following form
\be 
    \mathcal{O}_\eta(x) = \prod_{l =1}^{2Q} : e^{i\eta_l \phi(k_l) } : \, ,
    \label{eq:O_def}
 \ee
where $2Q$ is the number of Fermi points at point $x$. Here we label the Fermi points at position $x$ as follows:
\begin{itemize}
    \item for $x \in$~L  (then $2Q =2 $)
    \begin{equation}
        \begin{array}{clll}
         & k_1 := k_{{\rm L},1}(x) , & \quad & k_2 := k_{{\rm L},2}(x) , \\
         {\rm and } \qquad  & \phi(k_1) := \phi_{\rm L} ( k_{{\rm L},1}(x)) , &  &\phi(k_2) := \phi_{\rm L} (k_{{\rm L},2}(x) )
         \end{array}
    \end{equation}
        \item for $x \in$~R  (then $2Q =2 $)
    \begin{equation}
        \begin{array}{clll}
         & k_1 := k_{{\rm R},1}(x) , & \quad & k_2 := k_{{\rm R},2}(x) , \\
         {\rm and } \qquad  & \phi(k_1) := \phi_{\rm R} ( k_{{\rm R},1}(x)) , &  &\phi(k_2) := \phi_{\rm R} (k_{{\rm R},2}(x) )
         \end{array}
    \end{equation}
        \item for $x \in$ C  (then $2Q =4 $)
    \begin{equation}
        \begin{array}{llll}
        &k_1 := k_{{\rm L},1}(x) , \quad & k_2 := k_{{\rm R},1}(x) , \\
        &k_3 := k_{{\rm R},2}(x) , \quad & k_4 := k_{{\rm L},2}(x) , \\[4pt]
   {\rm and } \qquad & \phi(k_1) := \phi_{\rm L}(k_{{\rm L},1}(x)) ,  & \phi(k_2) := \phi_{\rm R}(k_{{\rm R},1}(x)) , \\
     & \phi(k_3) := \phi_{\rm R}(k_{{\rm R},2}(x)) , & \phi(k_4) := \phi_{\rm L}(k_{{\rm L},2}(x) ).
        \end{array}
    \end{equation}
\end{itemize}
The coefficients $\eta_l$ in the exponentials of Eq.~(\ref{eq:O_def}) can be interpreted as the displacement of the Fermi contour operated by the vertex operator $:e^{i\eta_l\phi(k_l)}:$ around the point $k_l$. Positive (negative) values of $\eta_l$ corresponds to inwards (outwards) deformations.
They are half-integers, $\eta_l \in \mathbb{Z} + \frac{1}{2}$, because the shifts correspond to a change of boundary conditions for the fermions from periodic to anti-periodic (see the next subsection for more on this). Moreover, the above bosonization formulas give
\begin{eqnarray}
    \left[ \int_{-\pi}^\pi ( \delta \rho_{\rm L}(k) + \delta \rho_{\rm R}(k)) dk , \mathcal{O}(x) \right] \, = \, - \left( \sum_{l=1}^{2Q} \eta_l \right)  \mathcal{O}_\eta(x) ,
\end{eqnarray}
and, since all operators $\mathcal{O}_\eta(x)$ appearing in the expansion (\ref{eq:exp_sigma}) must have charge $-1$ because they destroy one particle, we must have
\begin{equation}
    \label{eq:tot_charge}
    \sum_{l=1}^{2Q} \eta_l = 1.
\end{equation}
The scaling dimension of the operator (\ref{eq:O_def}) is
\begin{equation}
    \label{eq:tot_scaling}
    \Delta_\eta = \sum_{l=1}^{2Q} \frac{\eta_l^2}{2} .
\end{equation}
Looking at the sets of coefficients $\eta_l$ that are compatible with Eq.~(\ref{eq:tot_charge}) and minimize (\ref{eq:tot_scaling}), one finds the operators of minimal scaling dimension appearing in the expansion (\ref{eq:exp_sigma}).

For $2$ Fermi points ($Q = 1$) there is a single operator (\ref{eq:O_def}) of minimal scaling dimension, corresponding to
\be
\eta = (\eta_1, \eta_2)  = \left( + \frac{1}{2}, + \frac{1}{2} \right).
\ee
For $4$ Fermi points ($Q = 2$), there are four operators (\ref{eq:O_def}) of minimal scaling dimension corresponding to
\begin{equation}
    \begin{split}
        \eta = (\eta_1,\eta_2,\eta_3,\eta_4) =  \left( - \frac{1}{2}, + \frac{1}{2}, + \frac{1}{2}, + \frac{1}{2} \right), \\
        {\rm or } \qquad \eta = \left( + \frac{1}{2}, - \frac{1}{2}, + \frac{1}{2}, + \frac{1}{2} \right), \\
         {\rm or } \qquad \eta = \left( + \frac{1}{2}, + \frac{1}{2}, - \frac{1}{2}, + \frac{1}{2} \right) , \\
         {\rm or } \qquad \eta = \left( + \frac{1}{2}, + \frac{1}{2}, + \frac{1}{2}, - \frac{1}{2} \right).
         \label{eq:etas_Q2}
    \end{split}
\end{equation}
Now that we have identified the most relevant operators appearing in the right hand side of Eq.~(\ref{eq:exp_sigma}), let us turn to the coefficients $\mathcal{C}_\eta (x)$. Those coefficients include the Jacobian of the change of coordinates $x \mapsto k_{\alpha,a} (x)$  around each Fermi point ($\alpha = $L,R, $a=1,2$), and a numerical factor $B_\eta (\{ k_a \})$ which depends on the set of Fermi points at position $x$,
\begin{equation}
    \mathcal{C}_\eta(x) = B_\eta (\{ k_a (x)\}_{1 \leq a\leq 2Q} )\prod_{l=1}^{2Q} \left| \frac{\text{d}k_l }{\text{d}x} \right|^{1/8}.
\end{equation}
In Sec.~\ref{subsubsec:computation_B} below, we compute the numerical factors $B_\eta (\{ k_a \})$ by relating them to form factors, following Refs.~\cite{shashi2011nonuniversal,shashi2012exact,bondesan2015chiral,brun2018inhomogeneous,scopa2020one}. The result reads
\be
    \label{eq:formula_B}
B_\eta( \{ k_a \}_{1 \leq a \leq 2Q} ) \,  = \, 
     \frac{ \left( \frac{G^2(3/2)}{\sqrt{\pi}}\right)^Q }{\sqrt{2}}  \prod_{1 \leq a<b\leq 2Q} \left| 2 \sin ( \frac{k_b-k_a}{2} ) \right| ^{\eta_a \eta_b} , \qquad (Q \geq 1),
\ee
and $B_\eta = 0$ if $Q=0$, i.e. if there are no Fermi points at position $x$.

In summary, the final result of this paragraph is an expansion of the lattice boson creation/annihilation operators (\ref{eq:exp_sigma}) in terms of the field theory operators, which we can express as follows: 
\begin{align}
    \label{eq:exp_sigma_final1}
\nonumber \hat{\sigma}_x^+  = & \sum_{\eta \in \mathcal{I}}  \frac{ \left( \frac{G^2(3/2)}{\sqrt{\pi}}\right)^Q }{\sqrt{2}}  \prod_{1 \leq a<b\leq 2Q} \left| 2 \sin ( \frac{k_b-k_a}{2} ) \right| ^{\eta_a \eta_b}  \prod_{c=1}^{2Q} \left| \frac{\text{d}k_c }{\text{d}x} \right|^{1/8}   \prod_{l =1}^{2Q} : e^{-i\eta_l \phi(k_l) } : \,  \\
& \qquad \qquad  + \; {\rm less \; relevant \; operators},
\end{align}
\begin{align}
    \label{eq:exp_sigma_final2}
\nonumber \hat{\sigma}_x^-   = & \sum_{\eta \in \mathcal{I}}  \frac{ \left( \frac{G^2(3/2)}{\sqrt{\pi}}\right)^Q }{\sqrt{2}}  \prod_{1 \leq a<b\leq 2Q} \left| 2 \sin ( \frac{k_b-k_a}{2} ) \right| ^{\eta_a \eta_b}  \prod_{c=1}^{2Q} \left| \frac{\text{d}k_c }{\text{d}x} \right|^{1/8}   \prod_{l =1}^{2Q} : e^{i\eta_l \phi(k_l) } : \, \\
& \qquad \qquad  + \; {\rm less \; relevant \; operators} .
\end{align}
Here $2Q = 0,2,4$ is the number of Fermi points at position $x$, and the set of indices $\mathcal{I}$ in the sum (\ref{eq:exp_sigma}), we take
\begin{equation}
    \label{eq:indices_set}
    \begin{array}{lllcc}
     \mathcal{I} & := & \emptyset & {\rm if} & Q=0, \\
    \mathcal{I} & := & \{ (1/2,1/2) \}  & {\rm if} & Q=1, \\
    \mathcal{I} & := & \{ (-1/2,1/2,1/2,1/2), (1/2,-1/2,1/2,1/2), \\
    && \quad (1/2,1/2,-1/2,1/2), (1/2,1/2,1/2,-1/2) \} & {\rm if}&  Q=2.
    \end{array}
\end{equation}
Thus, the sums in Eqs. (\ref{eq:exp_sigma_final1})-(\ref{eq:exp_sigma_final2}) run over zero term (for $Q=0$), or one term (for $Q=1$) or four terms ($Q=2$). 
Eqs. (\ref{eq:exp_sigma_final1})-(\ref{eq:exp_sigma_final2}) are the key formulas that we use below to derive our main result for the OPDM.

\subsubsection{Computation of the numerical factor $B_\eta( \{k_a\} )$.}
\label{subsubsec:computation_B}

In this paragraph we explain how to derive Eq.~(\ref{eq:formula_B}). The numerical coefficients $B_\eta(\{ k_a\})$ can be extracted from the thermodynamic limit of the finite-size form factors of the operator $\hat{\sigma}^-_x$, as explained in detail in Refs.~\cite{bondesan2015chiral,brun2018inhomogeneous,scopa2020one,shashi2011nonuniversal,shashi2012exact}:
\begin{equation}
    \label{eq:form_factor}
    B_\eta (\{k_a\}_{1 \leq a \leq 2Q}) = \lim_{L \rightarrow \infty} \left( \frac{L}{2\pi} \right) ^{\Delta_\eta}
       \left| \bra{\{ q^\eta_i\}_{i=1}^{N-1}} \hat{\sigma}^- _x \ket{\{ p_j \}_{j = 1}^N} \right| ,
\end{equation}
for carefully chosen eigenstates $\left| \{ p_j \}_{j=1}^{N} \right>$ and $\left| \{ q^\eta_i \}_{i=1}^{N-1} \right>$ of the Hamiltonian (\ref{eq:XX}) for $L$ spins with periodic boundary conditions. For convenience, in this paragraph we label the sites of the chain by integers $x=0, \dots, L-1$ (as opposed half-integers in the rest of the paper). This convention can affect the phase of the matrix element $\bra{\{ q^\eta_i\}_{i=1}^{N-1}} \hat{\sigma}^- _x \ket{\{ p_j \}_{j = 1}^N}$, but since we are interested only in its modulus in Eq.~(\ref{eq:form_factor}), this does not change the result, and it simplifies the formulas below. Here the eigenstates are normalized, $\braket{ \{ p_j \}}{\{ p_j \}} = \braket{ \{ q_i^\eta \}}{\{ q^\eta_i \}} =1$.\\

All the eigenstates $\left| \{ p_j \} \right>$ are easily constructed using the fermionic form of the Hamiltonian (\ref{eq:hoppingH}): they are obtained by creating fermions with rapidities $p_a$ over the vacuum $\left| 0 \right>$,
\begin{equation}
    \left|  \{ p_j \}_{j=1}^N \right> \,: = \, \hat{c}^\dagger(p_1) \dots \hat{c}^\dagger(p_N) \left| 0 \right>,
\end{equation}
where $\hat{c}^\dagger (p) = \frac{1}{\sqrt{L}} \sum_{x = 0}^{L-1} e^{i p x} \hat{c}^\dagger_x$; notice that $\hat{c}^\dagger(p)$ has normalization $1/\sqrt{L}$ due to the finite size. Crucially, the set of allowed momenta depends on the parity of the number of bosons $N$,
\begin{equation}
    \label{eq:sets_even_odd}
    \begin{array}{ccl}
   N \; {\rm even}: &\qquad & p \in \{ -\frac{2\pi}{L}\frac{L-1}{2}, -\frac{2\pi}{L}\frac{L-3}{2} , \dots , \frac{2\pi}{L} \frac{L-3}{2}   , \frac{2\pi}{L}\frac{L-1}{2}  \} \\
    N \; {\rm odd}: &\qquad & p \in \{ -\frac{2\pi}{L}\frac{L}{2}, -\frac{2\pi}{L}\frac{L-2}{2} , \dots , \frac{2\pi}{L} \frac{L-4}{2} , \frac{2\pi}{L}\frac{L-2}{2}  \} ,
    \end{array}
\end{equation}
where we are assuming that the number of sites $L$ is even. This is a well-known consequence of the Jordan-Wigner transformation from the boson Hamiltonian (\ref{eq:XX}) to the free fermion one (\ref{eq:hoppingH}) for finite length $L$. While the bosons always obey  periodic boundary conditions (b.c.), the fermions have anti-periodic b.c. if $N$ is even, and periodic b.c. if $N$ is odd.

Now, let us assume that $N$ is odd. In formula (\ref{eq:form_factor}), a specific choice is made for the eigenstate $\left| \{ p_j \}_{j=1}^{N} \right>$ which depends on the set of Fermi momenta $\{ k_a \}_{1 \leq a \leq 2Q}$ in the l.h.s of (\ref{eq:form_factor}). This eigenstate is a finite-size realization of a `split Fermi sea' or `Moses state'~\cite{eliens2017quantum}, defined by the following set of rapidities:
\begin{equation}
    \begin{array}{c}
    \{ p_j \}_{j=1}^N \, := \,  ( [k_1,k_2] \cup [k_3,k_4] \cup \dots \cup [k_{2Q-1},k_{2Q}] )   \cap \{ -\frac{2\pi}{L}\frac{L}{2}, -\frac{2\pi}{L}\frac{L-2}{2} , \dots , \frac{2\pi}{L} \frac{L-4}{2} , \frac{2\pi}{L}\frac{L-2}{2}  \} .
    \end{array}
\end{equation}
Here we are simply assuming that this is consistent with $N$ being odd, and, for convenience, we define the number of fermions in each component of the split Fermi sea, $N_1, N_2,  \dots , N_Q$, such that
\begin{eqnarray}
    \label{eq:pjset}
 \nonumber   p_1, \dots , p_{N_1} &\in&  [k_1,k_2] , \\
  \nonumber      p_{N_1+1}, \dots , p_{N_2} &\in&  [k_3,k_4] , \\
 \nonumber      & \dots & \\
  \nonumber      p_{N_{Q-1}+1}, \dots , p_{N_Q} &\in&  [k_{2Q-1},k_{2Q}] . 
\end{eqnarray}
The other eigenstate in Eq.~(\ref{eq:form_factor}), $\left| \{ q^\eta_i \}_{i=1}^{N-1} \right>$, must contain $N-1$ particles, and this is an even number (because $N$ is assumed to be odd). Therefore its rapidities must live in the other set in Eq.~(\ref{eq:sets_even_odd}). A minimal way of constructing such an eigenstate starting from the set $\{ p_j\}_{j=1}^N$ is to select a sequence $\eta = ( \eta_a )_{1\leq a \leq 2Q}$ such that
\begin{equation}
    \eta_a = \pm \frac{1}{2} , \qquad \sum_{a=1}^{2Q} \eta_a = 1 ,
\end{equation}
and choose its set of rapidities as
\begin{align}
    \label{eq:qeta}
 \nonumber  
 &\quad \{ q^\eta_i \}_{i=1}^{N-1}    :=   \left\{ k_1 + \frac{2\pi \eta_1}{L} ,  k_1 + \frac{2\pi \eta_1}{L} +\frac{2\pi}{L} , \dots ,   k_2 - \frac{2\pi \eta_2}{L} \right\} \cup \\
\nonumber    
& \cup \left\{ k_3 + \frac{2\pi \eta_3}{L} , \dots,  k_4 - \frac{2\pi \eta_4}{L} \right\} \cup \dots \cup \left\{ k_{2Q-1} + \frac{2\pi \eta_{2Q-1}}{L} ,  \dots ,  k_{2Q} - \frac{2\pi \eta_{2Q}}{L} \right\} . \\
\end{align}
The idea is that the eigenstate $\ket{ \{q_i\}_{i=1}^{N-1} }$ obtained from this `minimal' modification of the rapidities in the eigenstate $\ket{ \{p_j\}_{j=1}^{N-1} }$ corresponds precisely to the lowest possible excitations around the split Fermi sea $\{k_a\}_{1\leq a \leq 2Q}$ in the thermodynamic limit $L \rightarrow \infty$. Formulas (\ref{eq:pjset})-(\ref{eq:qeta}) define the eigenstates entering in Eq. (\ref{eq:form_factor}).

Now let us turn to the matrix element 
$ \bra{\{ q^\eta_i\}_{i=1}^{N-1}} \hat{\sigma}^-_x \ket{\{ p_j \}_{j = 1}^N} $ in Eq.~(\ref{eq:form_factor}). The absolute value of that matrix element is independent of $x$ because both eigenstates are eigenvectors of the lattice translation operator. Therefore we can choose $x$ to be on the leftmost site of the chain ($x=0$ in our conventions) so that, under the Jordan-Wigner transformation, one simply has $\hat{\sigma}_{0}^- = \hat{c}_{0}$. Then the matrix element can be expressed as
\be
\bra{\{q^\eta_a\}_{a = 1}^{N-1}} \hat{\sigma}^-_{1/2} \ket{\{ p_b \}_{b=1}^N} = \bra{0}\hat{c} ( q^\eta_{N-1}) \dots \hat{c}(q^\eta_1) \hat{c}_{1/2} \hat{c}^\dagger(p_{1}) \dots \hat{c}^\dagger(p_{N}) \ket{0} .
\label{eq:matrixel}
\ee
Then the key step in the calculation is to observe that, for rapidities $p$ and $q$ belonging to the two different sets (\ref{eq:sets_even_odd}), the fermion annihilation operators are related through
\be
\hat{c}(p) = \frac{1}{L} \sum_{q} \frac{2}{2i \sin\left( \frac{p-q}{2} \right)} \hat{c}(q).
\ee
Then we use $\hat{c}_{0} = \frac{1}{\sqrt{L}} \sum_p  \hat{c}_p$, and we apply Wick's theorem to express the matrix element (\ref{eq:matrixel}) as a determinant,
\begin{equation}
 \bra{\{q^\eta_a\}_{a = 1}^{N-1}} \hat{\sigma}^-_{1/2} \ket{\{ p_b \}_{b=1}^N} = 
 \left| \begin{array}{cccc}
    \frac{1}{\sqrt{L}}  &\frac{2}{2Li \sin\left( \frac{p_1 - q^\eta_1}{2} \right)} &...& \frac{2}{2Li \sin\left( \frac{p_1 - q^\eta_{N-1}}{2} \right)} \\
   \frac{1}{\sqrt{L}}  &\frac{2}{2Li \sin\left( \frac{p_2 - q^\eta_1}{2} \right)} &...& \frac{2}{2Li \sin\left( \frac{p_2 - q^\eta_{N-1}}{2} \right)} \\
    \vdots  & \vdots & \ddots & \vdots \\
    \frac{1}{\sqrt{L}}  &\frac{2}{2Li \sin\left( \frac{p_N - q^\eta_1}{2} \right)} &...& \frac{2}{2Li \sin\left( \frac{p_N - q^\eta_{N-1}}{2} \right)}\\
\end{array}  \right| .
\end{equation}
The latter $N \times N$ determinant can be turned into a Cauchy determinant by adding a rapidity $q^\eta_0$ to the set $\{ q^\eta_{i} \}_{i=1}^{N-1}$, which one then sends to infinity as follows:
\begin{equation}
 \lim_{ q_0^\eta \rightarrow i \infty }  \;  \frac{\sqrt{L}}{2} e^{- i q_0^\eta/2} 
\left| \begin{array}{cccc}
     \frac{2}{2Li \sin\left( \frac{p_1 - q^\eta_0}{2} \right)}  &\frac{2}{2Li \sin\left( \frac{p_1 - q^\eta_1}{2} \right)} &...& \frac{2}{2Li \sin\left( \frac{p_1 - q^\eta_{N-1}}{2} \right)} \\
     \frac{2}{2Li \sin\left( \frac{p_2 - q^\eta_0}{2} \right)}  &\frac{2}{2Li \sin\left( \frac{p_2 - q^\eta_1}{2} \right)} &...& \frac{2}{2Li \sin\left( \frac{p_2 - q^\eta_{N-1}}{2} \right)} \\
    \vdots  & \vdots & \ddots & \vdots \\
    \frac{2}{2Li \sin\left( \frac{p_N - q^\eta_0}{2} \right)} &\frac{2}{2Li \sin\left( \frac{p_N - q^\eta_1}{2} \right)} &...& \frac{2}{2Li \sin\left( \frac{p_N - q^\eta_{N-1}}{2} \right)}\\
\end{array}  \right|  .
\end{equation}
Our matrix element is then
\begin{equation}
\label{eq:matrix2}
    \begin{split}
 \bra{\{q^\eta_a\}_{a = 1}^{N-1}} \hat{\Psi}(0) \ket{\{ p_b \}_{b=1}^N} \, =\,  \frac{\sqrt{L}}{2} \lim_{q \rightarrow i \infty} e^{-iq/2} \mathrm{det}\left( \left[ \frac{2}{2Li \sin\left( \frac{p_a - q^\eta_b}{2}\right)} \right]_{1\leq a \leq N, 0\leq b \leq N-1}\right)   \\
 \qquad \qquad = \, \frac{\sqrt{L}}{2} \lim_{q \rightarrow i \infty} e^{-iq/2}   \frac{2^N}{L^N} \frac{ \prod_{1 \leq a < b \leq N } 2i \sin\left( \frac{p_a - p_b}{2} \right) \prod_{1 \leq c < d \leq N } 2i \sin\left( \frac{q^\eta_a - q^\eta_b}{2} \right) }{\prod_{1 \leq i \leq N} \prod_{1 \leq j \leq N} 2i \sin\left( \frac{p_i - q^\eta_j}{2} \right) } ,
\end{split}
\end{equation}
leading to the final result for the modulus of the matrix element:
\begin{equation}
    \left| \bra{\{q^\eta_i\}_{i = 1}^{N-1}} \hat{\Psi}(0) \ket{\{ p_j \}_{j=1}^N } \right| = \frac{2^{N-1}}{L^{N-\frac{1}{2}}} \frac{\prod_{1 \leq a < b \leq N} \left| 2 \sin\left( \frac{p_a - p_b}{2} \right) \right| \prod_{1 \leq c < d \leq N-1} \left| 2 \sin\left( \frac{q^\eta_c - q^\eta_d}{2} \right) \right|}{\prod_{i = 1}^N \prod_{j = 1}^{N-1} \left| 2  \sin \left( \frac{p_i - q^\eta_j}{2} \right) \right| }.
\end{equation}
We then arrive at the following expression for the numerical factor $B_\eta (\{k_a\})$,
\begin{equation}
\label{eq:limit_form_factor}
    B_\eta (\{k_a\}_{1 \leq a \leq 2Q}) = \lim_{L \rightarrow \infty} \left( \frac{L}{2\pi} \right) ^{\frac{Q}{4}}
       \frac{2^{N-1}}{L^{N-\frac{1}{2}}} \frac{\prod_{1 \leq a < b \leq N} \left| 2 \sin\left( \frac{p_a - p_b}{2} \right) \right| \prod_{1 \leq c < d \leq N-1} \left| 2 \sin\left( \frac{q^\eta_c - q^\eta_d}{2} \right) \right|}{\prod_{i = 1}^N \prod_{j = 1}^{N-1} \left| 2  \sin \left( \frac{p_i - q^\eta_j}{2} \right) \right| } .
\end{equation}
It is this formula which we use to arrive at the final expression (\ref{eq:formula_B}) for the numerical factors. To evaluate the limit, we rely on the asymptotic formula (\ref{eq:conjecture2}) presented in detail in~\ref{app:conjecture}. This asymptotic formula directly leads to our final result (\ref{eq:formula_B}).

\subsection{Final result for the one-particle density matrix}

\subsubsection{Main steps in the calculation.}
Finally, we have all the ingredients that are needed to derive our result for the equal-time OPDM~(\ref{eq:result2})-(\ref{eq:h_def}). In particular, we have the expansion (\ref{eq:exp_sigma_final1})-(\ref{eq:exp_sigma_final2}) of the lattice boson creation/annihilation operators in terms of the field theory operators. We see that this expansion leads to the OPDM in the form of a vector-matrix-vector product,
\be 
g_1(x,y;t)  =\mathcal{C}^\dagger(x;t) \cdot F(x,y;t) \cdot \mathcal{C}(y;t),
\label{eq:vmv_product}
\ee
where the vector $\mathcal{C}(x;t)$ and the matrix $F(x,y;t)$ are of small finite dimension. Specifically, the vector $\mathcal{C}(x;t)$ is either the zero vector (in that case $g_1(x,y;t) = 0$), or it is of dimension $1$, or of dimension $4$, depending on the position $x$. This is best summarized in a table:
\begin{equation}
\begin{tabular}{c|c|c|c}
     Interval & Name & Dim. $\mathcal{C}(x;t)$ & $\mathcal{C}(x;t)$  \\ \hline \hline
     $x< -t$ or $x>N+t$  &  Empty (E)  &   & 0 \\ & & &  \\ 
     $-t < x < {\rm min}(t,N-t)$ &  Left (L) & 1 &  $\frac{G(3/2)^2}{\sqrt{\pi}} \left(  \frac{1}{(2t)^{1/4}} \right) $ \\ &&& \\
     $ N-t <x<t $ \quad (if $t>\frac{N}{2}$) &  Central (C) & 4 & $\frac{G(3/2)^4}{\sqrt{2} \pi} \begin{pmatrix}
    \frac{1}{(t^2 - x^2)^{1/4}} \\
    \frac{1}{(t^2 - x^2)^{1/4}} \\
    \frac{1}{(t^2 - (N-x)^2)^{1/4}} \\
    \frac{1}{(t^2 - (N-x)^2)^{1/4}} \\
     \end{pmatrix} $  \\ &&& \\
    $ {\rm max} (t,N-t) < x < N+t$ &  Right (R) & 1 & $ \frac{G(3/2)^2}{\sqrt{\pi}}  \left(\frac{1}{(2t)^{1/4}} \right)$  \\ &&& \\ 
     $ t<x< N-t$ \quad (if $t< \frac{N}{2}$) &  Filled (F)  &   & 0 .
\end{tabular}
\label{eq:regions}
\end{equation}
Those entries for the vector $\mathcal{C}$ are obtained by evaluating the coefficients in the expansion (\ref{eq:exp_sigma_final1})-(\ref{eq:exp_sigma_final2}), namely
\begin{equation}
    \mathcal{C}_\eta(x,t) = \frac{ \left( \frac{G^2(3/2)}{\sqrt{\pi}}\right)^Q }{\sqrt{2}}  \prod_{1 \leq a<b\leq 2Q} \left| 2 \sin ( \frac{k_b-k_a}{2} ) \right| ^{\eta_a \eta_b}  \prod_{c=1}^{2Q} \left| \frac{\text{d}k_c }{\text{d}x} \right|^{1/8} ,
\end{equation}
using the explicit form (\ref{eq:parameterization}) of $k_a(x)$.

Similarly, the matrix $F(x,y;t)$ is either the zero matrix or it is of size $1\times 1$, or $1\times 4$, or $4\times 1$ or $4 \times 4$ depending of the positions of $x$ and $y$:
\begin{equation}
    \begin{tabular}{c|ccccc}
        Dim. of $F(x,y;t)$  &  $x \in E$ & $x \in L$ & $x \in C $  & $x \in R$  & $x \in F$ \; \\ \hline \hline
        $y \in E$  &&&&& \\
        $y \in L$  && $1\times 1$ & $4 \times 1$ & $1 \times 1$ & \\
        $y \in C$  && $1\times 4$ & $4\times 4$ & $1\times 4$ &     \\
        $y \in R$  && $1\times 1$ & $4\times 1$ & $1\times 1$ &      \\
        $y \in F$   &&&&&      
    \end{tabular}
\end{equation}
and the void entries stand for the case where $F(x,y;t)$ is the zero matrix. Notice that the vector-matrix-vector product (\ref{eq:vmv_product}) is always well-defined. The entries of the matrix $F(x,y;t)$ are obtained by explicitly evaluating the correlation function of the field theory operators,
\begin{eqnarray}
    \label{eq:F_etaeta}
   F_{\eta^x, \eta^y} (x,y;t) &=&  \left< \prod_{l =1}^{2Q_x} : e^{-i\eta^x_l \phi(k_l(x)) }:   \prod_{m =1}^{2Q_y} : e^{i\eta^y_m \phi(k_m(y)) } : \right>
\end{eqnarray}
using Wick's theorem for the bosonic fields $\phi_{\rm R}$ and $\phi_{\rm L}$, along with its one- and two-point functions (\ref{eq:phi_correl_LR}). The expression (\ref{eq:F_etaeta}) is ambiguous because the field $\phi(k)$ is defined modulo $2\pi$, so the exponential $e^{\frac{i}{2}\phi}$ is defined only up to a sign. This sign is of course important for our purposes, since the OPDM $g_1(x,y;t)$ is a sum of such terms, which depends on those signs. We discuss how to fix this amibiguitity in full details in Sec.~\ref{subsec:phases} below.

This leads to the following entries:
{\small\begin{equation}
    \begin{tabular}{c|c|c|c}
        $F(x,y;t)$  & $x \in L$ &  $x \in C$ & $x \in R$ \\  \hline
     $y\in L$ &   $ M_1(x,y) $ & 
    $M_4(x,y)$
     & $ \left( 0 \right) $ \\ 
     $y \in C$ &  $M_4^\dagger (y,x)$
     &   $\left( \begin{array}{c|c}
        M_2 (x,y) \\ \hline
         &   M_2 (N-x,N-y) 
     \end{array} \right)$   & 
     $ M_4^\dagger(N-y,N-x) \cdot \left( \begin{array}{cc|cc}
                 &  & 1 &  \\
                 & & & 1 \\ \hline
                 1 & & & \\
                  & 1 & &
          \end{array} \right) $  \\
     $y \in R$ &  $\left( 0 \right)$  &
          $\left( \begin{array}{cc|cc}
                 &  & 1 &  \\
                 & & & 1 \\ \hline
                 1 & & & \\
                  & 1 & &
          \end{array} \right) \cdot M_4(N-x,N-y)$
     & $M_1(N-x,N-y)$
    \end{tabular}
    \label{eq:result3}
\end{equation}}
where the $1\times 1$ matrix $M_1$ is
\begin{equation}
    M_1 (x,y) \equiv  e^{i \frac{\pi}{2}(x-y)}  \frac{r(x)^{\frac{1}{4}} r(y)^{\frac{1}{4}}}{|x-y|^{\frac{1}{2}}} \, \times \, \left( 1 \right) ,
\end{equation}
the $4 \times 1$ matrix $M_4$ is ---we recall that $r(x)$, $\varphi(x)$, $D(x,y)$ are defined in Eq.~(\ref{eq:supp_funcs})---
\begin{equation}
M_4 (x,y) \equiv e^{i \frac{\pi}{2}(x-y)}  \frac{(t/2)^{\frac{1}{4}} \,r(x)^{\frac{1}{4}} r(y)^{\frac{1}{4}}}{ \, r(N-x)^{\frac{1}{4}} \, |x-y|^{\frac{1}{2}} }  \, \times \, \begin{pmatrix}
    0 \\
    0 \\
    e^{-i \varphi(N-x)}   \\
    e^{i \varphi(N-x)}
\end{pmatrix} ,
\end{equation}
and the $2\times 2$ matrix $M_2$ is 
\begin{equation}
\begin{split}
    M_2(x,y) \; \equiv \; & e^{-i \frac{\pi}{2} (x-y)}  \frac{(t/2)^{ \frac{1}{2}} \, r(N-x)^{\frac{1}{4}} r(N-y)^{\frac{1}{4}}}{r(x)^{\frac{1}{4}} r(y)^{\frac{1}{4}}|x-y|^{1/2}}  \times \\
    & \left( \begin{array}{cc}
            e^{-i \varphi(x)} \\ &  e^{i \varphi(x)}
        \end{array} \right) \left(
        \begin{array}{cc}
            D(x,y)^{-1/2} &  D(x,y)^{1/2}  \\
            D(x,y)^{1/2}  & D(x,y)^{-1/2} 
        \end{array}
    \right) \left( \begin{array}{cc}
            e^{i \varphi(y)} \\ &  e^{-i \varphi(y)}
        \end{array} \right) . 
\end{split}
\end{equation}
{
The cases $E$/$F$ are not present in eq.~\eqref{eq:result3} since $g_1(x,y)$ equals to zero if one of the two points $x,y$ lies in these intervals.} Evaluating the vector-matrix-vector product (\ref{eq:vmv_product}) explicitly, one then arrives at our main result (\ref{eq:result2})-(\ref{eq:h_def}).

\subsubsection{Comment on how to fix the phases. }
\label{subsec:phases}

Here we discuss how we fix the phase ambiguity in the calculation of the expectation value of the product of vertex operators
\begin{eqnarray}
    \left< \prod_{l =1}^{2Q_x} : e^{-i\eta^x_l \phi(k_l(x)) }:   \prod_{m =1}^{2Q_y} : e^{i\eta^y_m \phi(k_m(y)) } : \right> ,
\end{eqnarray}
which defines the matrix elements of the matrix $F(x,y;t)$. This expression involves operators on the left and right contours, but it is clear that it factorizes into two independent contributions, one from each contour. For instance, let us focus on the case when all operators are on the same contour, e.g. when $x,y \in L$. In that case we need to evaluate a four-point function
\begin{eqnarray}
    \left<  : e^{-\frac{i}{2} \phi(k_{{\rm L}, 1}(x) ) }: \,  : e^{-\frac{i}{2} \phi(k_{{\rm L}, 2}(x) ) }: \,  : e^{\frac{i}{2} \phi(k_{{\rm L}, 1}(y) ) }: \,  : e^{\frac{i}{2} \phi(k_{{\rm L}, 2}(y) ) }:    \right> .
    \label{eq:vertex_corr}
\end{eqnarray}
This quantity is ill-defined because the field $\phi(k)$ is only defined modulo $2 \pi$. However, the quantity $\partial_k \phi(k)$ is always well defined, it measures the fluctuations of the density of HCBs along the Fermi contours, as seen in Eqs.~(\ref{eq:densityfluct1},\ref{eq:densityfluct2}).
So we define the correlation function (\ref{eq:vertex_corr}) as 
\begin{equation}
    \left<  e^{- \frac{i}{2} \int_{\Gamma} \partial_k \phi(k) \, dk  } \right> ,
\end{equation}
properly regulated at short distance, where $\Gamma$ is a union of oriented paths along the contour that go from $k_{{\rm L},1}(x)$ to $k_{{\rm L},1}(y)$ and from $k_{{\rm L},2}(x)$ to $k_{{\rm L},2}(y)$, see the figure in the top-left entry of Table~\ref{table:contours}. The choice of the oriented path $\Gamma$ fixes the phase ambiguity in the correlation (\ref{eq:vertex_corr}). Once the phase ambiguity has been fixed by selecting a set of oriented paths, the correlation function is evaluated using standard methods of bosonization, leading to the following expression for  Eq.~(\ref{eq:vertex_corr})
\begin{eqnarray}
\nonumber
   && \left<  e^{- \frac{i}{2} \int_{\Gamma} \partial_k \phi(k) \, dk  } \right> \, = \, e^{-\frac{i}{2} \int_{\Gamma} \partial_q \Phi(q) dq}  e^{-\frac{1}{8} \int_{\Gamma} dq \int_{\Gamma} dq' \left< \partial_q \phi(q) \partial_{q'}\phi(q') \right>_{\rm conn.}} \\ && =\, e^{-\frac{i}{2} \int_{\Gamma} \partial_q \Phi(q) dq} \left| \frac{2 \sin \left(\frac{k_{{\rm R},1}(x) - k_{{\rm R},2}(x)}{2}  \right) 2 \sin \left(\frac{k_{{\rm L},1}(y) - k_{{\rm L},2}(y)}{2} \right)}{\prod_{i,j = 1,2} 
    2 \sin \left(\frac{k_{{\rm L},i}(x) - k_{{\rm R},j}(y)}{2} \right)} \right|^{1/4} .
    \label{eq:vertex_int}
\end{eqnarray}

More generally, in Table \ref{table:contours} we show our conventions for the choice of the oriented path $\Gamma$ for all different cases. 
The signs $+, -$ stand for $\pm \frac{1}{2}$ in the exponent in the vertex operators associated to their given points along the contour, and the oriented path is drawn in black. We always choose our oriented paths so that they go from a `$-$' to `$+$'. We also choose our convention so that, when four points are along the same contour, the associated phase factor is one when the two points $x$ and $y$ coincide. In the case of two points on the same contour, the conventions are the following: the associated phase factor disappears if the point is at $N-t$ on the right contour and at $t$ on the left contour (notice that these conventions are consistent with the symmetries of the system and the formulas shown in Eqs~(\ref{eq:result2} - \ref{eq:supp_funcs})).\\

The general expression that we use for the correlation function (\ref{eq:F_etaeta}) is:

\begin{multline}
    \left<  e^{- \frac{i}{2} \int_{\Gamma} \partial_k \phi(k) \, dk  } \right>  = e^{-\frac{i}{2} \int_{\Gamma} \partial_q \Phi(q) dq} \times
    \\ 
      \frac{\prod \limits_{\alpha = {R,L}}  \prod \limits_{1 \leq a<b \leq Q_x} \left| 2 \sin \left( \frac{k_{\alpha,{\rm a}}(x) - k_{\alpha,{\rm b}}(x)}{2} \right) \right|^{1/4} \prod \limits_{\beta = {R,L}} \prod \limits_{1 \leq c<d \leq Q_y} \left| 2 \sin\left( \frac{k_{\beta, {\rm c}}(y) - k_{\beta,{\rm d}}(y)}{2} \right) \right|^{1/4} }{  \prod \limits_{\gamma = {R,L}} \prod \limits_{l=1}^{Q_x} \prod \limits_{m=1}^{Q_y} \left| 2 \sin\left( \frac{k_{\gamma,{\rm a}}(x) - k_{\gamma,{\rm m}}(y)}{2} \right) \right| ^{1/4} } .
    \label{eq:vertex_int_general}
\end{multline}

\begin{table}[h!]
    \begin{tabular}{c||c|c|c|}
        $\Gamma(x,y;t)$  & $x \in L$ &  $x \in C$ & $x \in R$ \\  \hline\hline
     $y\in L$ &   \includegraphics[width = 0.22\textwidth]{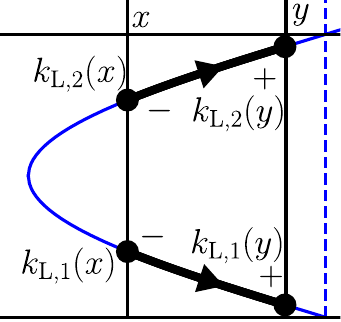} & 
    $ \includegraphics[width = 0.22\textwidth]{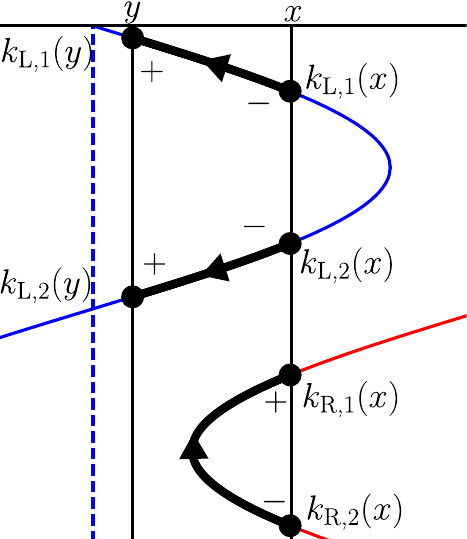}$
     & $ $ \\ \hline \hline
     
     $y \in C$ & \includegraphics[width = 0.22\textwidth]{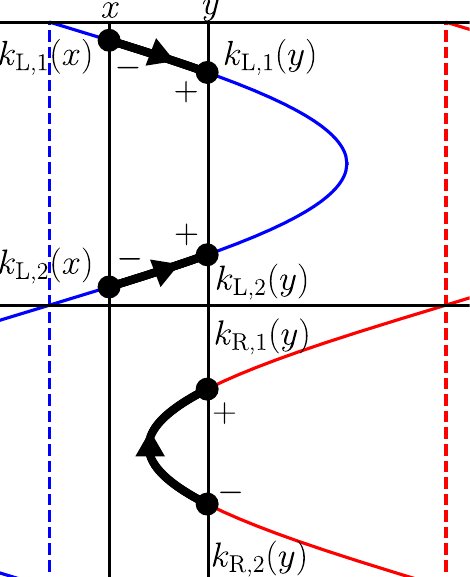}
     &    \includegraphics[width = 0.22\textwidth]{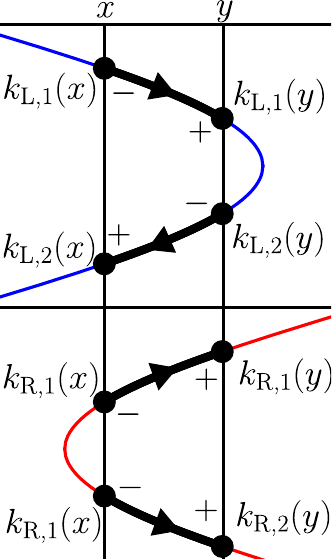}   & 
      \includegraphics[width = 0.22\textwidth]{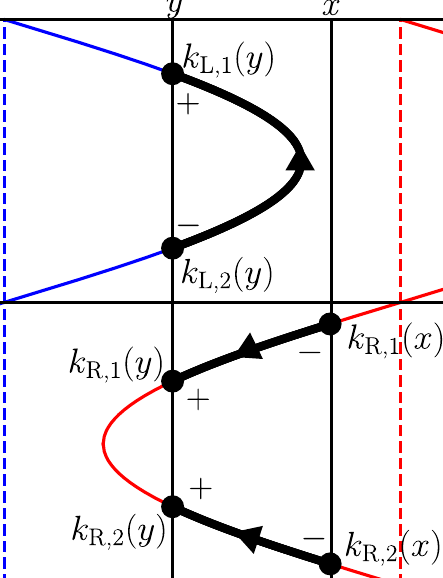}\\ \hline\hline
      
     $y \in R$ &    &
           \includegraphics[width = 0.22\textwidth]{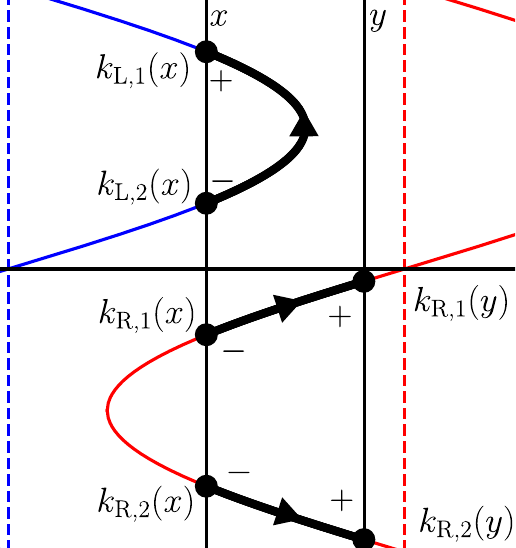}
     &  \includegraphics[width = 0.22\textwidth]{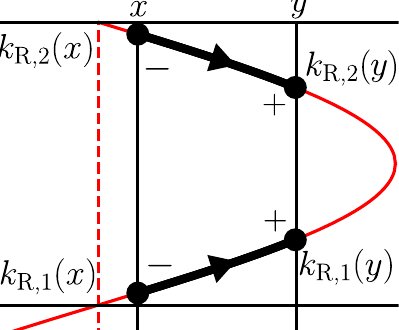} 
    \end{tabular}
    \caption{Our conventions for the choice of the oriented path $\Gamma$ for all different cases. The signs $+, -$ stand for $\pm \frac{1}{2}$ in the exponent in the vertex operators associated to their given points along the contour, and the oriented path is drawn in black, with the arrows denoting the orientation. }
    \label{table:contours}
\end{table}


\begin{figure}[t!]
    \includegraphics[width=\textwidth]{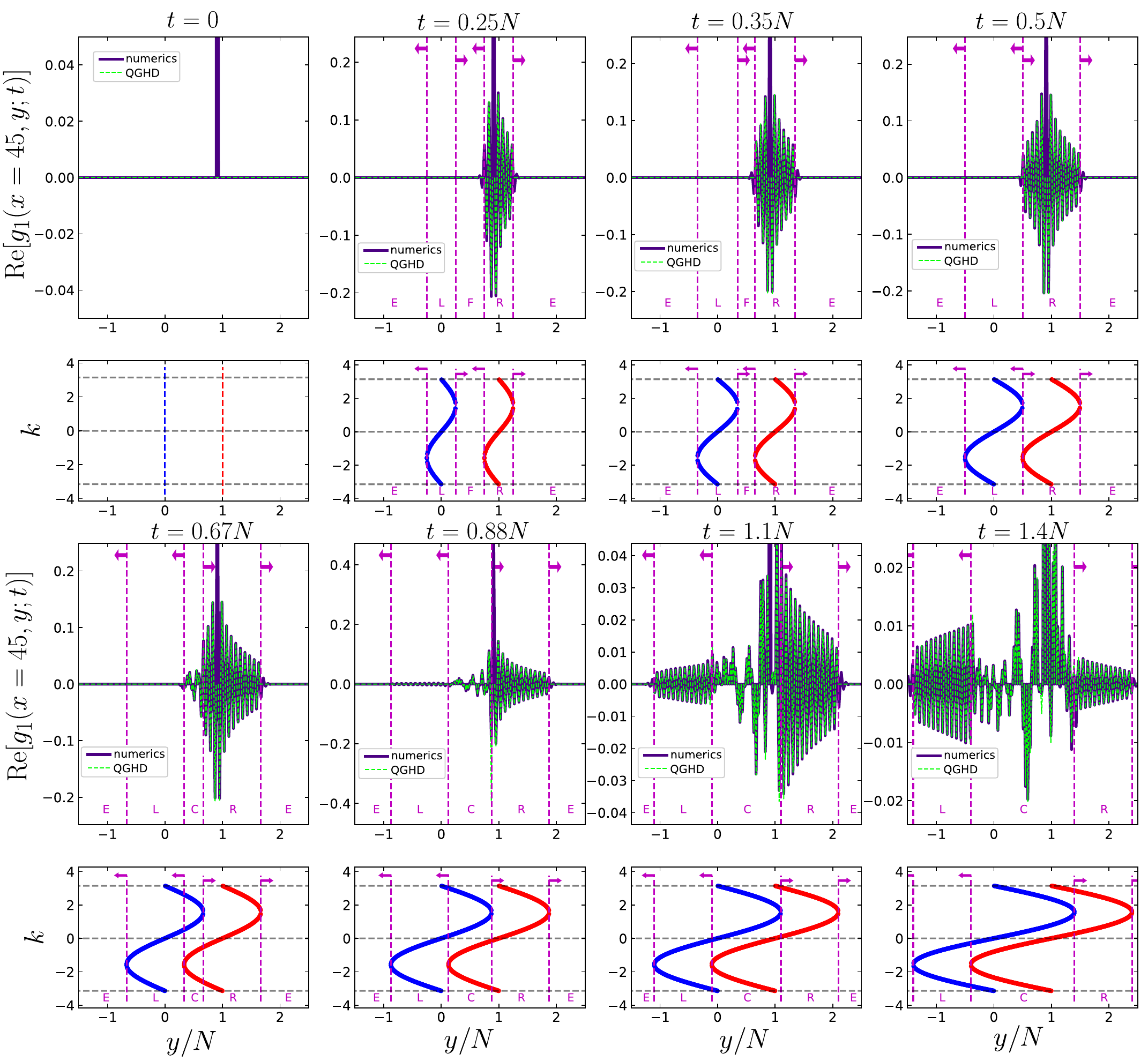}
    \caption{Real part of $g_1(x,y;t)$ as a function of $y/N$ for a fixed value $x = 45$, and particle number $N = 50$. {
    The upper panels show ${\rm Re}[g_1(x=45,y;t)]$ and the lower panels show} the associated Fermi contours $\Gamma$ (blue~-~left contour, red~-~right contour) at different times, showing the dynamical development of long-range correlations in the system as well as the appearance of the regions with split Fermi seas. The dashed purple vertical lines denote the boundaries of the different regions, the purple letters denote the different regions with the notation introduced in Eqs~(\ref{eq:regions0}) - (\ref{eq:result2}) and used throughout the paper (e.g., in Eq.~(\ref{eq:result3})). The arrows show how the region boundaries move with time. Grey dashed horizontal lines mark the points $-\pi$, $0$ and $\pi$. Analytical results (solid lines) and exact numerical calculations (dashed lines) are found in perfect agreement at all times.} \label{fig:ODLRO}
\end{figure}

 \section{Quasicondensation and  development of off-diagonal long-range order}
\label{sec:applications}

In this section, we make use of our formula for the OPDM to discuss the phenomena of quasicondensation and dynamical fermionization, occurring at different stages during the free expansion of the hard-core gas. 

To begin with, it is instructive to show the evolution of the OPDM \eqref{eq:OPDM} for a fixed number of bosons $N$ and position $x$, see Fig.~\ref{fig:ODLRO}. 
This figure illustrates the emergence of real-space correlations during the free propagation of HCBs and its connection to the development of off-diagonal long-range order in the OPDM. 
At $t=0$ there are no correlations as the initial state is a product state.
At early times $t<N/2$, non-vanishing correlations are only observed within the `right front' region (R) ---see the nomenclature in the table in Eq.~\eqref{eq:regions0}--- and are associated with the spreading of the front around $x=N$. 
Within this region, whose size  grows linearly in time up to $t\sim N/2$, the qualitative properties of the OPDM remain unchanged. 
At time $t=N/2$, the fastest particles originating from the left front (at $x=0$) meet the ones from the right front, and a new --qualitatively different-- central region (C) with non-vanishing correlations emerges. 
This is characterized by the presence of a split Fermi sea arising from the intersection of the two light cones corresponding to each of the two Fermi contours. 
Interestingly, upon the convergence of particles coming from the two domain walls, the system around position $x$ establishes non-vanishing correlations also with the regions (L), $|y|/t<1$, where particles propagating from the right interface have not arrived yet, underscoring the entangled nature of the HCBs. The OPDM \eqref{eq:result2} displays the same power-law decay at large distance  $g_1\propto |x-y|^{-1/2}$ for $x, y \in $ R (or $x,y \in $ L), for almost the entire dynamics, as noticed numerically in Ref.~\cite{Rigol2004quasicond}.

At large times $t\gg N$, the central region (C) grows and eventually governs the entire bulk behavior. 
In this regime, $x,y \in \text{C}$, the system behaves as a homogeneous gas of length $\ell =2t-N$ containing two species of entangled particles (one from each junction), with long-range spatial correlations decaying as $g_1 \sim |x-y|^{-1}$ when $t\gg N$; the half-system entanglement saturates to the value $S \sim \frac{1}{3}\log(N/2)$ when $t\gg N$ \cite{Eisler_2021,scopa2022exact}. 
As we shall see below, the different power-law decay of spatial correlations is physically associated with the quasicondensation ($g_1\propto |x-y|^{-1/2}$) or the dynamical fermionization ($g_1\propto |x-y|^{-1}$) of the HCBs, respectively.
Our exact formula \eqref{eq:result2} characterizes both these regimes, as well as the crossover occurring between the two. 

Fig.~\ref{fig:ODLRO}  offers a qualitative depiction of the development of off-diagonal long-range order. 
A more detailed and quantitative examination of quasicondensation is elaborated in the following two subsections.

\begin{figure}[t!]
    \centering
    \includegraphics[width=\textwidth]{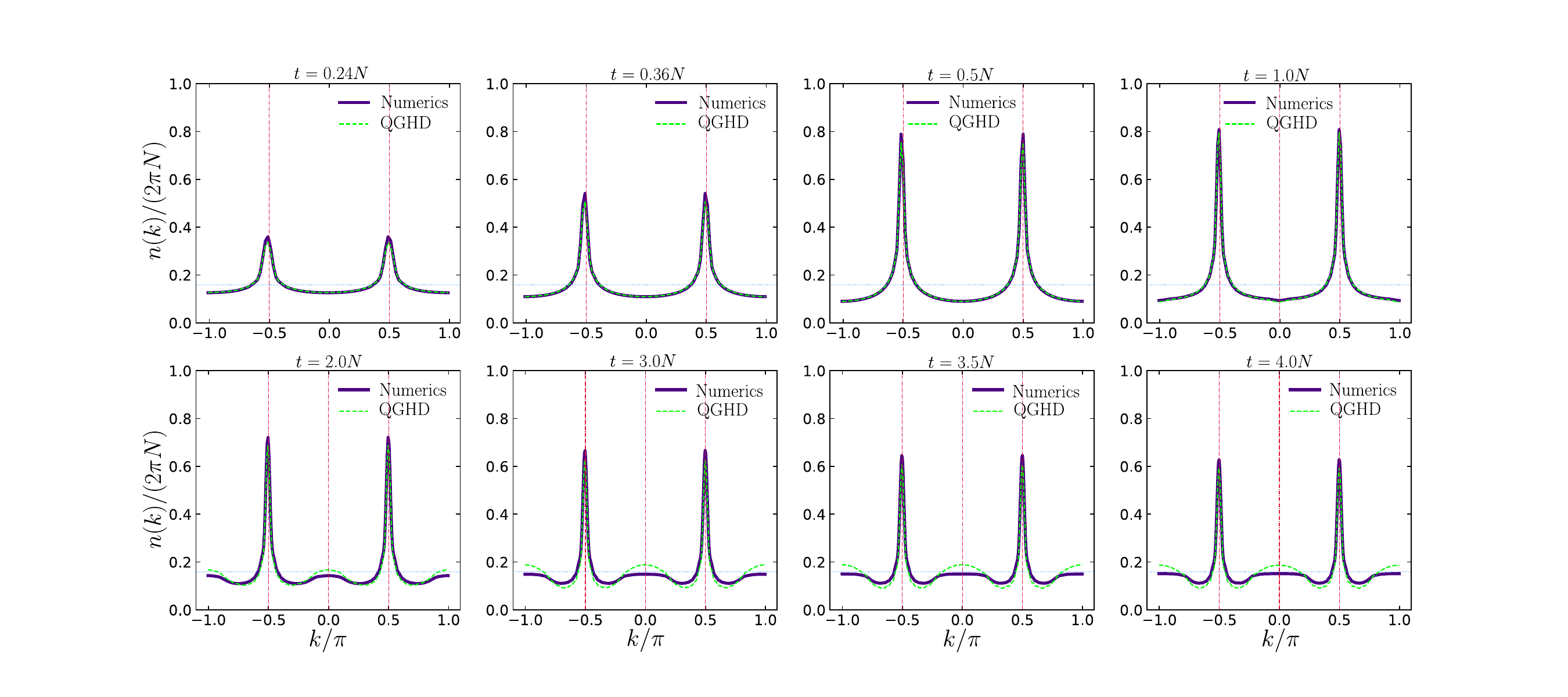}
    \caption{Momentum distribution function for the same setting as in Fig.~\ref{fig:ODLRO}. The two peaks centered at $p=\pm \pi/2$ correspond to the fastest excitations and serve as a signal quasicondensation of the HCBs. 
    The horizontal dotted line shows the asymptotic value of the momentum distribution during the free expansion, $n(p)=N$, as expected by dynamical fermionization. 
    The data obtained by Fourier transforming the (UV-regularized) analytical results for the OPDM (solid lines) and the exact numerical calculations (dashed lines) are found in perfect agreement at all times.}
    \label{fig:MD}
\end{figure}
\subsection{Evolution of the momentum distribution}
We first discuss the time evolution of the momentum distribution
{
\be
n(p; t)=\sum_x \sum_y \; e^{ip(x-y)} g_1(x,y;t),
\ee}
which numerically can be obtained via fast Fourier transform of the OPDM data. 
Here the normalization of the  momentum distribution is such that {
$\sum_p  n(p;t)=2\pi N$, $|p|\leq \pi$.}
We recall that our result for the equal-time OPDM is only valid at large distances, i.e. when $|x-y|\gg {\rm max}(n^{-1}(x),n^{-1}(y))$, 
because it is based on a field theory calculation. 
Hence it does not reproduce correctly the behavior of the OPDM at short distances. 
A UV-regularization is therefore necessary before taking the Fourier transform of Eq.~\eqref{eq:result2} to get the momentum distribution. 
We find that, for times $t/N\lesssim 1$, it is sufficient to regularize the value for $g_1(x,y;t)$ at coincident points $x= y$.  In other words, we replace the diagonal of the OPDM with the local value of density, i.e. we set $g_1(x,x;t)=n(x;t)$.
We report the resulting evolution of the momentum distribution in Fig.~\ref{fig:MD} (for the times corresponding to the snapshots of Fig.~\ref{fig:ODLRO}). 
The employed UV-regularization is good enough to perfectly match the numerical curve up to time $t/N\sim 1$. 
At later times $t/N \gg 1$, some deviations are observed. 
These discrepancies can be mitigated by systematically incorporating higher terms in the short-distance expansion of the OPDM, see e.g. Refs.~\cite{olshanii2003short,bouchoule2021breakdown}. 
Delving into improved UV-regularization schemes is beyond the scope of this paper, particularly since nearly perfect agreement with numerics is achieved by solely regularizing the diagonal of $g_1(x,y; t\lesssim N)$.

Our formula for the equal-time OPDM \eqref{eq:result2}  quantitatively captures the development of off-diagonal long-range order. Specifically,  throughout the majority of the dynamics, $g_1\propto |x-y|^{-1/2}$ for $x, y \in$ R (or $x,y \in $ L)  up to subleading corrections. 
This power-law decay is reflected also in the appearance of two peaks in the momentum distribution located at $p=\pm \pi/2$. 
At large times $t/N \gg 1$, the behavior of the OPDM changes to $g_1\propto |x-y|^{-1}$ (c.f. Eq.~\eqref{eq:result2}), and the peaks in the momentum distribution are expected to disappear ($n(p) \rightarrow N$ when $ t/N\gg 1$ corresponding to dynamical fermionization). 
However, showing this trend in the momentum distribution proves challenging due to finite-size corrections impacting numerical calculations at large times and the necessity for refined UV-regularizations for our analytical result.
Nonetheless, the plots in the second row of Fig.~\ref{fig:MD} already show a clear onset of dynamical fermionization at large times.

\subsection{Quasicondensation: scaling of the eigenvalues of $g_1(x,y;t)$}

Finally, after having shown the development of off-diagonal long-range order in the equal-time OPDM, we will now delve into its connection with the phenomenon of quasicondensation. 
To accomplish this, we examine the eigenvalues $\lambda_i(t)$ of the equal-time OPDM, i.e.
{
\be
\sum_x \ g_1(x,y;t)\ \xi_i(y)= \lambda_i \ \xi_i(y),
\ee}
with eigenvectors $\xi_i$ of the OPDM known as the `natural orbitals'. The scaling of the largest eigenvalue $\lambda_0$ of $g_1(x,y;t)$ with the number of atoms,
\begin{equation}
    \lambda_0 \propto N^\mathcal{C},
    \label{eq:largesteigval}
\end{equation}
defines an exponent ${\cal C}$ which gives information about the highest occupied natural orbital.{
We stress that in Eq.~(\ref{eq:largesteigval}) we are looking at the scaling with $N$ for a fixed ratio $t/N$.}
According to the Penrose-Onsager criterion \cite{Penrose-Onsager56}, an exponent ${\cal C}$ equal to one, i.e. $\lambda_0\propto N$, signals a condensate, because a finite fraction of all the bosons occupies the same natural orbital.
However, in a 1D with charge conservation, ${\cal C}$ is always strictly smaller than one because of the Mermin-Wagner theorem.
Nevertheless, as long as ${\cal C}$ is close enough to one, it is customary to say that the system is a quasicondensate— a phase which in modern literature extends all the way up to ${\cal C}=1/2$ \cite{lenard1964momentum,Yukalov_2005,forrester2003finite,colcelli2018universal,Rigol2004quasicond}.
In contrast, in non-interacting fermion systems, the fermion OPDM would give ${\cal C}=0$ because of Pauli exclusion and, expectedly, there is no quasicondensate. 

 \begin{figure}[t!]
\centering
    \includegraphics[width=.9\textwidth]{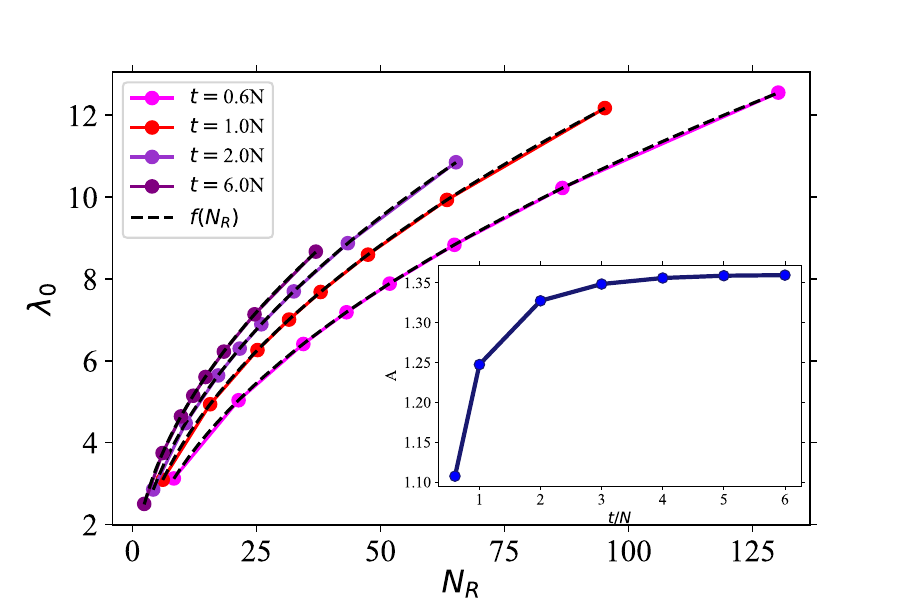}
    \caption{{
    Largest eigenvalue of the OPDM $\lambda_0$ plotted as a function of $N_R$ (symbols) at different times $t/N$. Solid lines and markers show the numerical values of $\lambda_0$ as a function of $N_R$ for different ratios $t/N$, while the black dashed lines show the function $f(N_R)= A N_R^{1/2} + B$, with amplitude $A$ and additive constant $B$ used as fit parameters. Inset: the fitted amplitude $A$ saturates for large $t/N$.}  }
    \label{fig:largesteigval}
\end{figure}

Our result for the OPDM \eqref{eq:result2} shows the presence of off-diagonal long-range order $g_1(x,y;t)\propto |x-y|^{-1/2}$ that persists for almost the entire evolution {
in the R and L regions. 
Such behaviour suggests to use the following scaling ansatz for the (normalized) natural orbital corresponding to the largest eigenvalue:
\be\label{eq:NO-scaling1}
\xi_0(x) \sim N_R^{-1/2}\  \Xi_0(x/N_R).
\ee
Here $N_R$ is the number of atoms in the R region, which scales as
\begin{equation}
    N_R \simeq \frac{N^{3/2}} {t^{1/2}}
\end{equation}
for large $t$, as follows from integrating the density profile (\ref{eq:densityprofile}) over the region R. We consider the Ansatz (\ref{eq:NO-scaling1}) for a fixed ratio $t/N$. An analogous Ansatz $\xi_0(x) \sim N^{-1/2}\  \Xi_0(x/N)$ was also tested numerically in Ref.~\cite{Rigol2004quasicond}. Replacing the sums in $\xi_0$
by integrals for $N_R\gg 1$, one can evaluate the largest eigenvalue (for fixed ratio $t/N$) as
\begin{align}\label{eq:NO-scaling2}
\lambda_0&= \sum_x \sum_y \ \xi^*_0(x) \ g_1(x,y;t) \ \xi_0(y)\propto \int_R dx \int_R dy \frac{\xi^*_0(x) \xi_0(y)}{|x-y|^{1/2}}\\
\nonumber
&\quad\propto\sqrt{N_R} \int dX \int dY \frac{\Xi^*_0(X) \Xi_0(Y)}{|X-Y|^{1/2}}\propto \sqrt{N_R},
\end{align}}
i.e. ${\cal C}= 1/2$, as previously noted in Refs.~\cite{lenard1964momentum,Yukalov_2005,forrester2003finite,colcelli2018universal,Rigol2004quasicond}. {
We have confirmed this scaling numerically using our formula for the OPDM, see Fig.~\ref{fig:largesteigval}. We find that, for fixed ratio $t/N$, the largest eigenvalue scales as $\lambda_0 \simeq A \sqrt{N_R} + B$ for large $N_R$, with an amplitude $A$ that saturates for large $t/N$.}

\begin{figure}[ht]
  \centering
  (a) \hspace{5cm} (b)\\
  \begin{minipage}{.495\textwidth}
 \includegraphics[width=.9\textwidth]{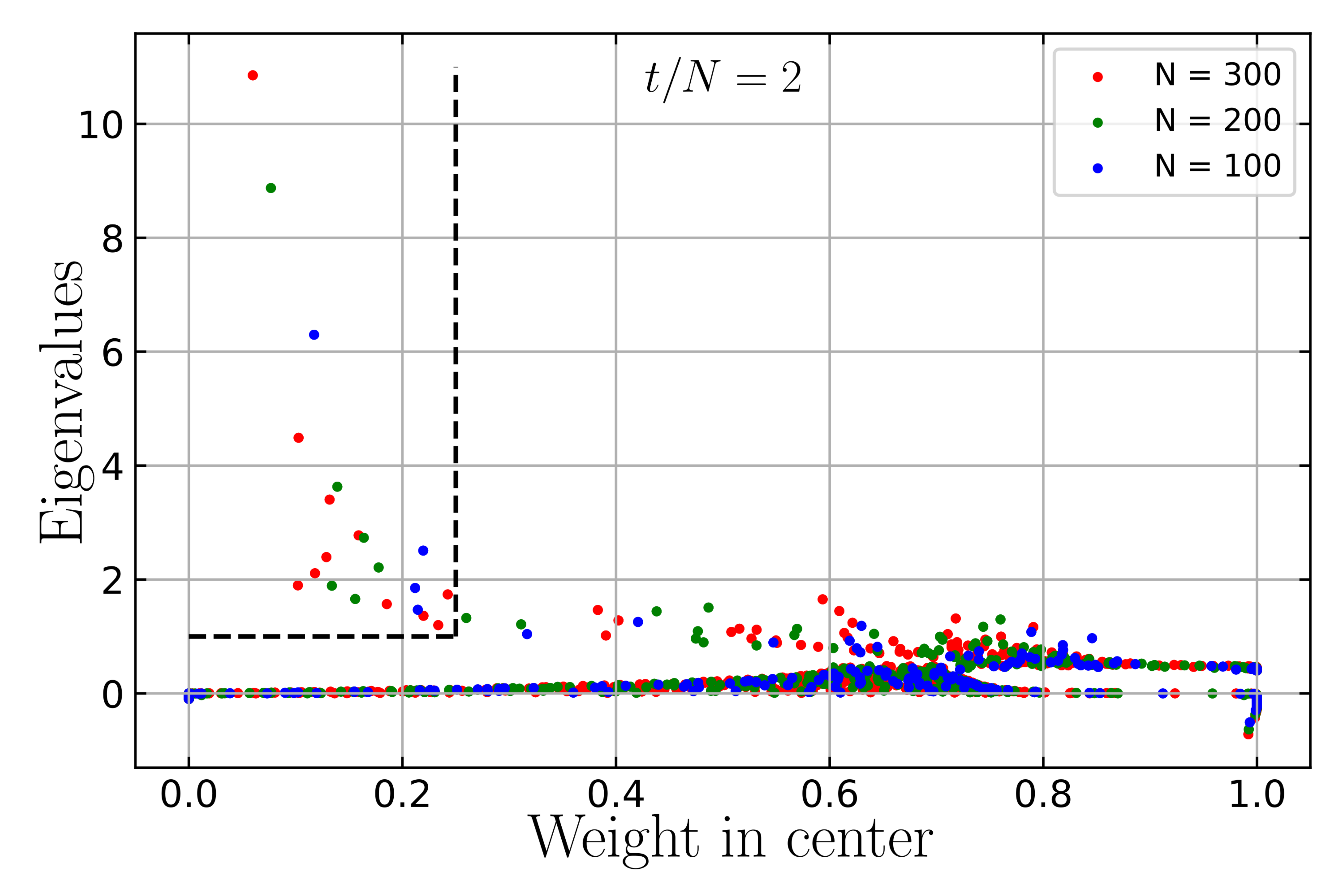}
 \end{minipage}
 \begin{minipage}{.495\textwidth}
 \includegraphics[width=.9\textwidth]{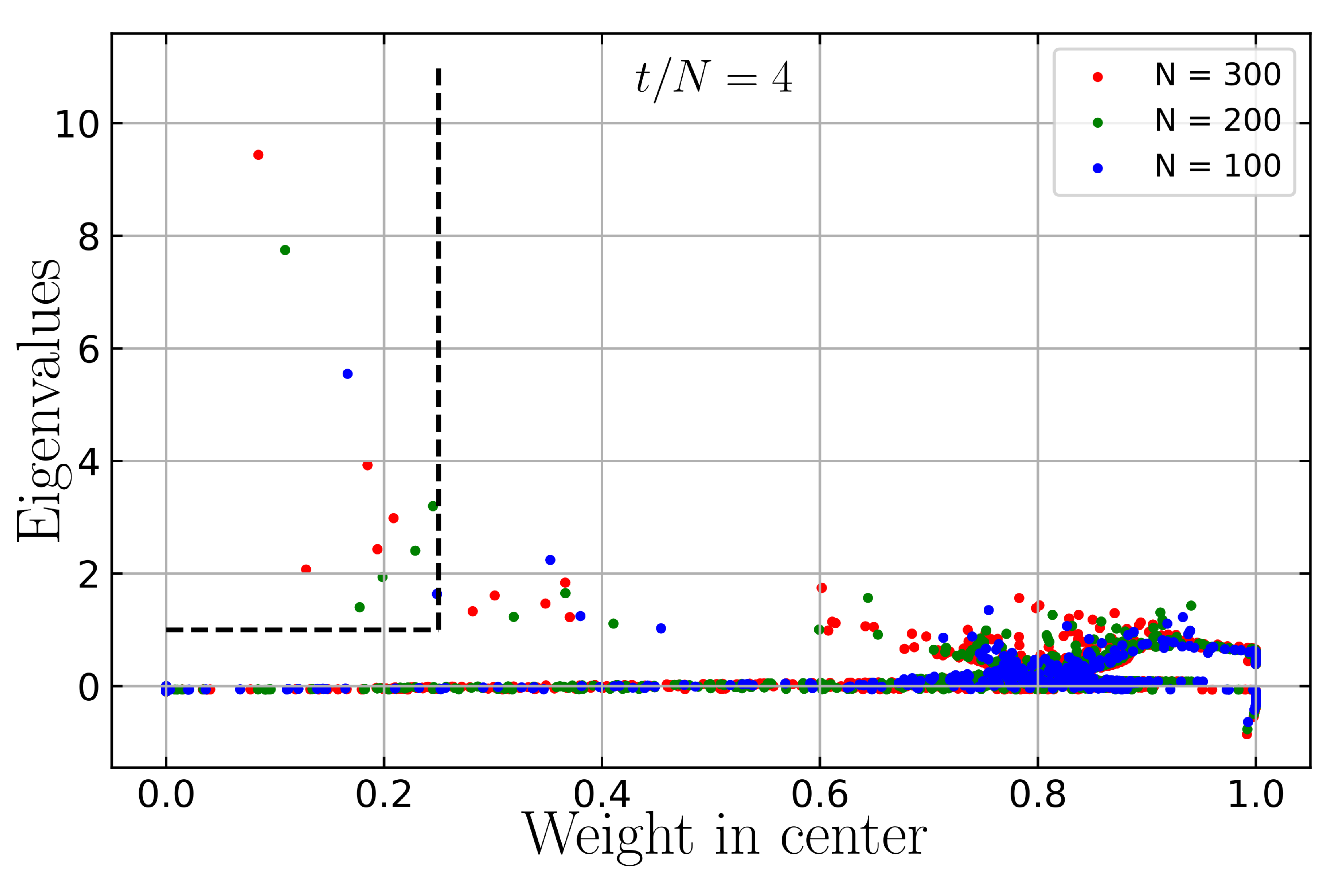}
 \end{minipage}
  \caption{Eigenvalues of the OPDM as a function of their weight in region (C) (see Eq.~\eqref{eq:weight}) for different $N$ and fixed $t/N$: (a) $t/N = 2$, (b) $t/N = 4$.  Since our analytical results are obtained in the thermodynamic limit, each eigenvalue is doubly degenerate (i.e. each point in the figure correspond to two eigenvalues), reflecting a perfect left/right symmetry of our setup. At finite size, an hybridization of the L/R degenerate eigenstates is observed, see Ref.~\cite{Rigol2004quasicond}. Dashed boxes enclose the eigenvalues with weight $w_i\leq 0.25$, see Fig.~\ref{fig:percentofstates}.}
 \label{fig:eigval_weight}
\end{figure}

Conversely, at large times $t/N\gg 1$,  we would expect an exponent ${\cal C}\approx 0$ as a signal of the emergent dynamical fermionization. {
However, this is seemingly in contradiction with the scaling shown in Fig.~\ref{fig:MD}, where the exponent seems to remain equal to $1/2$ even at times $t/N\gg 1$ for which the onset of dynamical fermionization is visible in the momentum distribution (c.f.~Fig.~\ref{fig:MD}).
This seeming contradiction is understood by noticing that the scaling shown in Fig.~\ref{fig:largesteigval} implies $\lambda_0 \sim A \sqrt{N_R} \sim A  (N/t)^{1/4} \sqrt{N}$, so while the exponent is always $1/2$, the  amplitude $A  (N/t)^{1/4}$ of that algebraic scaling slowly moves toward zero as time grows.}

\begin{figure}[ht]
    \centering
    \includegraphics[width=.65\textwidth]{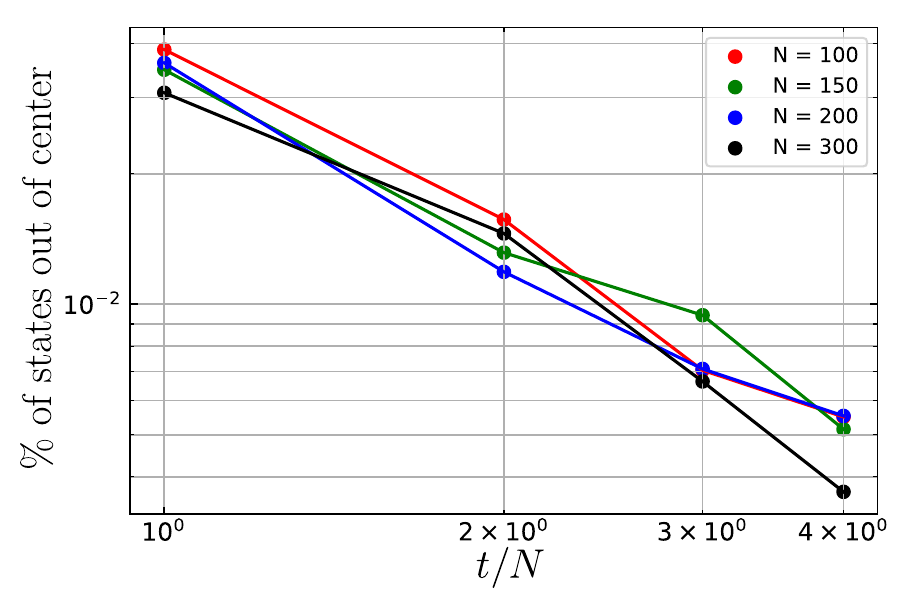}
    \caption{Percentage of eigenstates having most of their weight outside of the central region, $w_i\leq 0.25$, as a function of $t/N$ for different system sizes. The figure shows a clear decay of this quantity with time. In Fig.~\ref{fig:eigval_weight}, the dashed box encloses the eigenvalues having $w_i\leq 0.25$. }
    \label{fig:percentofstates}
\end{figure}

The dynamical fermionization is further clarified by examining the complete spectrum of $g_1(x,y,t)$ and its corresponding eigenstates.
For each eigenstate $\xi_i$ of the OPDM, we define the quantity
\be\label{eq:weight}
w_i=\int_{x\in \text{C}} dx \ |\xi_i(x)|^2,
 \ee
which assigns a weight $w_i$ to $\lambda_i$ according to the localization in real space of the natural orbital. The closer $w_i$ is to 1, the more the eigenstate is localized in the central region (C), and vice versa. In Fig.~\ref{fig:eigval_weight} we show this quantity for two different times. 
We see that the largest eigenstates have $w_i\simeq 0$, hence they are localized outside the central region, precisely in the (L/R) lobes of the system. Their eigenvalues scale as $N^{1/2}$, and they are responsible for the peaks at $\pm \pi /2$ in the momentum distribution (c.f.~Fig.~\ref{fig:MD}). Conversely, as we move to lower eigenvalues in the spectrum of $g_1$, we find a large number of eigenstates localized in the central region, with eigenvalues scaling as $N^0$. 
These groups of lower eigenstates are in fact those responsible for dynamical fermionization. As time flows, on the one hand the group of largest eigenstates will be supported on outer regions that are pushed to infinity, on the other hand, the fraction of states supported in the bulk increase and dictate the onset of dynamical fermionization, see Fig.~\ref{fig:percentofstates}. We remark that, since the regions (L) and (R) have a size that is independent on $t$ (whereas the central region grows linearly with $t$) and since the particle density $n(x)\sim 1/t$ when $t/N \gg 1$ (see \cite{scopa2022exact} and Eq.~\eqref{eq:density}), the group of largest eigenstates have population decreasing as $1/t$, although they would disappear only at time strictly equal $\infty$. 
At large but finite $t/N\gg 1$, their presence explains the value of the fitted exponent ${\cal C}\approx 1/2$ in Fig.~\ref{fig:largesteigval}.

In conclusion, our analysis show that there is a group of eigenvalues with scaling $\mathcal{C} =\frac{1}{2}$ and another with $\mathcal{C}=0$, the latter becoming more dominant with time, eventually causing dynamical fermionization.

\section{Summary and conclusion}\label{sec:conclusion}

In this work we studied the dynamics of the OPDM of 1D HCBs resulting from the releasing of a domain wall state--see Fig.~\ref{fig:lattice}(a) for an illustration.
Exploiting the recent development of quantum generalized hydrodynamics,
we derived an exact analytic formula in the Euler scaling limit (i.e., large space and time scales, $x, N, t\to \infty$ at fixed $x/t$, $N/t$). 
Utilizing this formula, we provided an analytical understanding of the phenomena associated with dynamic quasicondensation and fermionization, manifesting at different stages of the time evolution of the gas.
Moreover, our result gave access also to the intermediate times connecting the two stages.
We also performed exact numerical calculations, and we compared them to our prediction for all regimes of the dynamics and for different values of $N$ (see, e.g., Figs.~\ref{fig:OPDM_result} and \ref{fig:OPDM_result2}). 
We observed a perfect matching, with our result on top of the numerical data.

We stress that Eq.~\eqref{eq:result2} is one of very few available formulas for the equal-time OPDM of trapped HCBs that is explicitly written in terms of simple functions.
(Other known example concerns, e.g., the dynamics in harmonic traps, see \cite{forrester2003finite,minguzzi05,collura2013equilibration,collura2013quench,ruggiero2019conformal}.)
The derivation of a such explicit formula relies on both the choice of initial state and to the strong repulsion limit that has been considered. Nonetheless, its calculation is non-trivial, and thus
we find its final expression in Sec.~\ref{sec:summary} remarkably compact and simple.
Furthermore, the results in this manuscript lay the groundwork for determining the OPDM in various non-equilibrium scenarios involving one-dimensional hard-core bosons. 

Looking ahead, we plan to extend these computations to Bose gases with finite interactions. This extension demands special attention due to the loss of conformal invariance in such systems, see, e.g., the derivations in Refs.~\cite{brun2018inhomogeneous,scopa2020one,ruggiero2020quantum} for other observables.

\vspace{0.6cm}

{\it Acknowledgements}. We thank Paola Ruggiero for discussions and joint work on closely related topics {
and the anonymous Referee for useful comments about the scaling analysis of Sec. 4.2.} \\
 We acknowledge support from `Lorraine Université d'Excellence' program (AT), the ERC Consolidator Grant 771536 (NEMO) (SS and PC), ERC Starting Grant 101042293 (HEPIQ) (SS), from Agence
Nationale de la Recherche through ANR-20-CE30-0017-02 project ‘QUADY’ (JD) and ANR-22-CE30-0004-01 project
‘UNIOPEN’ (JD), and the Slovenian Research and Innovation Agency (ARIS), Research core funding Grants No.~P1-0044, N1-0273 and J1-50005 (LV).

\appendix
\section*{\Large Appendices}

\section{Two conjectures about asymptotic limits of products used in Sec.~\ref{subsubsec:computation_B}}
\label{app:conjecture}

In the main text we use the asymptotic formula Eq.~(\ref{eq:conjecture2}) to compute the numerical factors $B_\eta(\{k_a\})$, see paragraph~\ref{subsubsec:computation_B}. In this Appendix, we explain where the asymptotic formulas (\ref{eq:conjecture1}) and (\ref{eq:conjecture2}) come from. We have not been able to prove these two formulas beyond the special case $Q=1$, but we have performed extensive numerical checks, and we provide a physical intuition based on electrostatics that strongly supports them.

\paragraph{Conjecture 1.}

Let $K_1 < K_2 < \dots < K_{2Q}$ be integers, and $\eta_j = \pm \frac{1}{2}$ for $j=1,\dots, 2Q$. Let us consider the set of integers
\begin{equation*}
    \mathcal{I} := \mathbb{Z} \cap ([K_1, K_2] \cup [K_3, K_4] \cup \dots \cup [K_{2Q-1},K_{2Q}])
\end{equation*}
and the set of half-integers
\begin{equation*}
    \mathcal{J} := (\mathbb{Z} + \frac{1}{2}) \cap ([K_1 + \eta_1, K_2 - \eta_2] \cup [K_3+\eta_3, K_4-\eta_4] \cup \dots \cup [K_{2Q-1}+\eta_{2Q-1},K_{2Q}-\eta_{2Q}]) .
\end{equation*}
The cardinals of these two sets are $|\mathcal{I}| = \sum_{j=1}^{2Q} (K_{2j}-K_{2j-1}+1)$ and $|\mathcal{J}| = \sum_{j=1}^{2Q} (K_{2j}-K_{2j-1}+1-\eta_{2j}-\eta_{2j-1})$.

Then we have the following asymptotic formula when all the $K_j$'s are large:
\begin{equation}
    \label{eq:conjecture1}
    \frac{\prod_{a<b \in \mathcal{I}} |b-a|  \prod_{c<d \in \mathcal{J}} |d-c| }{ \prod_{i \in \mathcal{I}, j \in \mathcal{J}} |i-j| } \, \simeq \, \left( \frac{G(3/2)}{\pi^{1/4}} \right)^{2Q} \sqrt{\pi}^{|\mathcal{I}| + |\mathcal{J}| } \prod_{1 \leq a<b \leq 2Q} |K_b - K_a|^{\eta_a \eta_b} .
\end{equation}
Here the notation `$\simeq$' means that the ratio of the l.h.s and the r.h.s goes to one in the limit $|K_1|, |K_2|, \dots ,|K_{2Q}| \rightarrow \infty$, keeping the ratios $K_2/K_1$, $K_3/K_1$ etc. fixed.

\paragraph{Conjecture 2.}

Let $K_1 < K_2 < \dots < K_{2Q} < L$ be integers, and $\eta_j = \pm \frac{1}{2}$ for $j=1,\dots, 2Q$. Let $\mathcal{I}$ and $\mathcal{J}$ be defined exactly as in Conjecture 1.

Then we have the following asymptotic formula when all the $K_j$'s and $L$ are large:
\begin{eqnarray}
    \label{eq:conjecture2}
\nonumber    && \frac{\prod_{a<b \in \mathcal{I}} |\frac{L}{\pi} \sin \frac{\pi (b-a) }{L}|  \prod_{c<d \in \mathcal{J}} |\frac{L}{\pi} \sin \frac{\pi (d-c) }{L}|  }{ \prod_{i \in \mathcal{I}, j \in \mathcal{J}} |\frac{L}{\pi} \sin \frac{\pi (j-i) }{L}|  } \\
    && \simeq \, \left( \frac{G(3/2)}{\pi^{1/4}} \right)^{2Q} \sqrt{\pi}^{|\mathcal{I}| + |\mathcal{J}| } \prod_{1 \leq a<b \leq 2Q} \left|\frac{L}{\pi}  \sin \frac{\pi( K_b - K_a)}{L} \right|^{\eta_a \eta_b} .
\end{eqnarray}

\paragraph{Physical argument supporting the two conjectures.} Let us start with the first conjecture. The  product is equal to the 2D Coulomb energy of a set of charges placed along a line:
\begin{equation}
     \frac{\prod_{a<b \in \mathcal{I}} |b-a|  \prod_{c<d \in \mathcal{J}} |d-c| }{ \prod_{i \in \mathcal{I}, j \in \mathcal{J}} |i-j| } = \exp (-E) 
\end{equation}
where
\begin{eqnarray*}     
     && E = -\sum_{a<b \in \mathcal{I} }\log |a-b| - \sum_{c<d \in \mathcal{I} }\log |c-d| + \sum_{i\in \mathcal{I}, j \in \mathcal{J}} \log|i-j|  \\
     && =  \frac{1}{2} \fint dx \, dy  \, (\rho_{\mathcal{I}}(x) - \rho_{\mathcal{J}}(x) ) (\rho_{\mathcal{I}}(y) - \rho_{\mathcal{J}}(y) )  (- \log |x-y| ),
\end{eqnarray*}
with $\rho_\mathcal{I} (x) := \sum_{i \in \mathcal{I}} \delta(x-i)$ and $\rho_\mathcal{J} (x) := \sum_{j \in \mathcal{J}} \delta(x-j)$. The integral needs to be properly regulated when $x=y$, hence the notation `$\fint$'. Now, the point is that in the limit of large $K_j$'s, we may view $\rho_\mathcal{I}(x)$ and $\rho_\mathcal{J}(x)$ as continuous densities of charges. Inside all intervals $[K_j, K_{j+1}]$ these continuous densities are the same, except at the boundary points where they differ by $\pm \eta_j$. Namely, we have
\begin{equation}
    \rho_\mathcal{I} (x) - \rho_\mathcal{J} (x) \, \simeq \, \sum_{j=1}^{2Q} \eta_j \delta(x-K_j) .
\end{equation}
In other words, at large scale the system of `positive' charges in $\mathcal{I}$ and `negative' charges in $\mathcal{J}$ looks neutral, except at the boundary points $K_j$ where there is a `residual' charge $\eta_j$. Thus, the Coulomb energy of the full set of charges is dominated by the Coulomb energy of those residual charges. Moreover, because of the short-distance regularisation of the integral when $x=y$, we expect to have a first subleading  contribution to the Coulomb energy proportional to the length of the intervals, then a second subleading correction of order $O(1)$ associated with each boundary point, and then corrections that go to zero in the limit of large $K_j$'s. In summary, in that limit we expect to have
\begin{equation}
    E \, = \, \sum_{1\leq i<j \leq 2Q} - \eta_i \eta_j \log |K_i-K_j| \, + \,  D_0 \frac{ |\mathcal{I}| \,  + \, |\mathcal{J}|  }{2} \, + \,  \sum_{j=1}^{2Q} D_{\eta_j} \, + \, o(1)
\end{equation}
where $D_0$, $D_{1/2}$ and $D_{-1/2}$ are three unknown constants, and $o(1)$ are subleading terms that go to zero. Here we choose to write the total length of the intervals as $(|\mathcal{I}| + |\mathcal{J}|)/2$, but one could write it differently, for instance simply as $|\mathcal{I}|$, or as $|\mathcal{J}|$. These different choices would lead to differences of terms of order $O(1)$ that can be absorbed in a redefinition of the constants $D_{1/2}$ and $D_{-1/2}$.

This physical argument then leads to the conclusion that the product should behave as
\begin{equation}
    \label{eq:phys_conj1}
     \frac{\prod_{a<b \in \mathcal{I}} |b-a|  \prod_{c<d \in \mathcal{J}} |d-c| }{ \prod_{i \in \mathcal{I}, j \in \mathcal{J}} |i-j| } \simeq \prod_{1 \leq i<j \leq 2Q} |K_i-K_j|^{\eta_i \eta_j} \, (e^{-D_0/2 })^{|\mathcal{I}| + |\mathcal{J}|} \, \prod_{j=1}^{2Q} e^{-D_{\eta_j}} .
\end{equation}
The constants $D_0$, $D_{1/2}$ and $D_{-1/2}$ still need to be fixed. But, importantly, these constants are independent of the $K_j$'s and $\eta_j$'s.

Now let us turn to the second conjecture. The product is still equal to the 2D electrostatic energy of a distribution of charges in the plane, but this time along a circle of diameter $L/\pi$,
\begin{equation}
  \frac{\prod_{a<b \in \mathcal{I}} |\frac{L}{\pi} \sin \frac{\pi (b-a) }{L}|  \prod_{c<d \in \mathcal{J}} |\frac{L}{\pi} \sin \frac{\pi (d-c) }{L}|  }{ \prod_{i \in \mathcal{I}, j \in \mathcal{J}} |\frac{L}{\pi} \sin \frac{\pi (j-i) }{L}|  }  = e^{-E'}
\end{equation}
with 
\begin{equation}
    E' = - \sum_{a < b \in \mathcal{I}} \log |z_a- z_b| - \sum_{c < d \in \mathcal{J}} \log |z_c- z_d| + \sum_{i \in \mathcal{I},j \in \mathcal{J}} \log |z_i- z_j| ,
\end{equation}
where we use the following notation,
\begin{equation}
    z_n \, = \, \frac{L}{2\pi} e^{i \frac{2\pi}{L} n} \in \mathbb{C} .
\end{equation}
Then, by the same argument as above, in the limit of large $K_j$'s and $L$, keeping all ratios of length fixed, we expect the Coulomb energy to be dominated by the residual charges at the endpoints of the support of the charge distribution. We also expect to have a subleading extensive term proportional to the total length $(|\mathcal{I}|+|\mathcal{J}|)/2$ and a term of order $O(1)$ for each endpoint. The latter two terms should involve the same constants $D_0$, $D_{1/2}$, $D_{-1/2}$ as in conjecture 1, because they are local contribution, insensitive to the global geometry of the distribution of the charges. In conclusion, we expect
\begin{equation}
    E' \, = \, \sum_{1\leq i<j \leq 2Q} - \eta_i \eta_j \log |z_{K_i}-z_{K_j}| \, + \,  D_0 \frac{ |\mathcal{I}| \,  + \, |\mathcal{J}|  }{2} \, + \,  \sum_{j=1}^{2Q} D_{\eta_j} \, + \, o(1)
\end{equation}
and thus
\begin{eqnarray}
\nonumber  && \frac{\prod_{a<b \in \mathcal{I}} |\frac{L}{\pi} \sin \frac{\pi (b-a) }{L}|  \prod_{c<d \in \mathcal{J}} |\frac{L}{\pi} \sin \frac{\pi (d-c) }{L}|  }{ \prod_{i \in \mathcal{I}, j \in \mathcal{J}} |\frac{L}{\pi} \sin \frac{\pi (j-i) }{L}|  }  \\
  && \simeq  \prod_{1\leq i,j \leq 2Q} \left| \frac{L}{\pi} \sin \frac{\pi (K_i-K_j)}{L} \right|^{\eta_i \eta_j} (e^{-D_0/2 })^{|\mathcal{I}| + |\mathcal{J}|} \, \prod_{j=1}^{2Q} e^{-D_{\eta_j}}
\end{eqnarray}
with the same constants $D_0$, $D_{1/2}$ and $D_{-1/2}$ as above.

\paragraph{Fixing the constants $D_0$, $D_{1/2}$, $D_{-1/2}$.} It turns out that the case of a single interval, i.e. $Q=1$, is easier, and we can do the calculation exactly. This allows to identify the constants $D_0$, $D_{1/2}$, $D_{-1/2}$. We focus first on the case
\begin{equation}
    \mathcal{I} = \{ 1, \dots , N \} \quad {\rm and} \quad \mathcal{J} = \{ 1+1/2, \dots , N -1/2 \},
\end{equation}
corresponding to $\eta_1 = \eta_2 = 1/2$.
In that special case, one finds the following exact formula for any $N$:
\begin{equation}
    \label{eq:exact_barnes}
      \frac{\prod_{a<b \in \mathcal{I}} |b-a|  \prod_{c<d \in \mathcal{J}} |d-c| }{ \prod_{i \in \mathcal{I}, j \in \mathcal{J}} |i-j| } =  \pi^{N-1} G(3/2)^2 \frac{G(N) G(N+1)}{G(N+1/2)^2} .
\end{equation}
Here $G(.)$ is the Barnes $G$-function. Using the asymptotics of the Barnes $G$-function, one finds
\begin{equation}
    \frac{G(N) G(N+1)}{G(N+1/2)^2} = N^{1/4} + O(N^{-3/4})
\end{equation}
for large $N$. Then comparing Eq.~(\ref{eq:exact_barnes}) to Eq.~(\ref{eq:phys_conj1}), using $(|\mathcal{I}|+|\mathcal{J}|)/2 = N-1/2$, one finds that $e^{-D_0} = \pi$ and $e^{-D_{1/2}} = \frac{G(3/2)}{\pi^{1/4}}$.
To fix the constant $D_{-1/2}$, we focus on the case
\begin{equation}
    \mathcal{I} = \{ 1, \dots , N \} \quad {\rm and} \quad \mathcal{J} = \{ 1+1/2, \dots , N + 1/2 \} ,
\end{equation}
corresponding to $\eta_1 = 1/2$ and $\eta_2 = -1/2$. In that case we have the exact formula for any $N$:
\begin{equation}
      \frac{\prod_{a<b \in \mathcal{I}} |b-a|  \prod_{c<d \in \mathcal{J}} |d-c| }{ \prod_{i \in \mathcal{I}, j \in \mathcal{J}} |i-j| } = \pi^{N-1/2}  G(3/2)^2 \frac{\Gamma(N)}{\Gamma (N+1/2)} \frac{G(N) G(N+1)}{G(N+1/2)^2}  .
\end{equation}
The asymptotics of the Gamma function gives $\Gamma(N) / \Gamma (N+1/2) =  N^{-1/2} + O(N^{-3/2})$.
Then, comparing this with Eq.~(\ref{eq:phys_conj1}), and using $(|\mathcal{I}|+|\mathcal{J}|)/2=N$, we find that $e^{-D_{-1/2}} = \frac{G(3/2)}{\pi^{1/4}}$, which is the same value as for the constant $e^{-D_{1/2}}$. This then leads to the two conjectures (\ref{eq:conjecture1}) and (\ref{eq:conjecture2}) above.

\section*{\Large References}\rhead{}

\begin{thebibliography}{99}

\bibitem{De_Nardis_2018}
De Nardis J and Panfil M 2018,
\href{http://dx.doi.org/10.1103/PhysRevLett.120.217206}{{\em Phys. Rev. Lett.} {\bf 120} 21 }


\bibitem{Bocini2023}
Bocini S 2023,
\href{http://dx.doi.org/10.21468/SciPostPhys.15.1.027}{{\em SciPost Physics\/} {\bf 15} 1 }

\bibitem{greiner_mandel_02b}
Greiner M, Mandel O, H\"ansch T~W and Bloch I 2002, \href{https://www.nature.com/articles/415039a}{{\em Nature\/} {\bf 419} 51}.

\bibitem{kinoshita2006quantum}
Kinoshita T, Wenger T and Weiss D~S 2006, \href{https://www.nature.com/articles/nature04693}{{\em Nature\/} {\bf 440} 900}.

\bibitem{Calabrese_2016}
Calabrese P, Essler F~H~L and Mussardo G 2016, \href{https://doi.org/10.1088%2F1742-5468%2F2016%2F06%2F064001}{{\em J. Stat. Mech.} {\bf {(2016)}} 064001}.


\bibitem{Caux_2016}
Caux J~S 2016, \href{https://doi.org/10.1088%2F1742-5468%2F2016%2F06%2F064006}{{\em J. Stat. Mech.}{\bf { (2016)}} 064006}.
  

\bibitem{Essler_2016}
Essler F~H~L and Fagotti M 2016, \href{https://doi.org/10.1088%2F1742-5468%2F2016%2F06%2F064002}{
{\em J. Stat. Mech.} {\bf (2016)} 064002}.

\bibitem{cazalilla2011one}
Cazalilla M, Citro R, Giamarchi T, Orignac E and Rigol M 2011, \href{https://journals.aps.org/rmp/abstract/10.1103/RevModPhys.83.1405}{{\em
  Rev.~Mod.~Phys.\/} {\bf 83} 1405}.

\bibitem{rigol2004universal}
Rigol M and Muramatsu A 2004, \href{https://journals.aps.org/pra/abstract/10.1103/PhysRevA.70.031603}{ {\em Phys. Rev. A} {\bf 70} 031603}.

\bibitem{girardeau2000breakdown}
Girardeau M and Wright E 2000, \href{https://journals.aps.org/prl/abstract/10.1103/PhysRevLett.84.5239}{ {\em Phys. Rev. Lett.} {\bf 84} 5239}.

\bibitem{vignolo2013universal}
Vignolo P and Minguzzi A 2013, \href{https://journals.aps.org/prl/abstract/10.1103/PhysRevLett.110.020403}{ {\em Phys. Rev. Lett.} {\bf 110} 020403}.

\bibitem{wright_rigol_14}
Wright T~M, Rigol M, Davis M~J and Kheruntsyan K~V 2014, \href{https://link.aps.org/doi/10.1103/PhysRevLett.113.050601}
{ {\em Phys. Rev.
  Lett.} {\bf 113} 050601}.
  
\bibitem{rigol2007relaxation}
Rigol M, Dunjko V, Yurovsky V and Olshanii M 2007, \href{https://link.aps.org/doi/10.1103/PhysRevLett.98.050405}{ {\em Phys. Rev. Lett.} {\bf 98} 050405}.

\bibitem{rigol_muramatsu_06}
Rigol M, Muramatsu A and Olshanii M 2006, \href{https://link.aps.org/doi/10.1103/PhysRevA.74.053616}{{\em Phys. Rev. A\/} {\bf 74}
  053616}.

\bibitem{Vidmar_2016}
Vidmar L and Rigol M 2016,\href{https://doi.org/10.1088%2F1742-5468%2F2016%2F06%2F064007}{{\em J. Stat. Mech.} {\bf (2016)} 064007}.


\bibitem{Rigol2004quasicond}
Rigol M and Muramatsu A 2004, \href{{http://dx.doi.org/10.1103/PhysRevLett.93.230404}}{{\em Phys. Rev. Lett.} {\bf 93} 230404}.

\bibitem{vidmar_iyer_17}
Vidmar L, Iyer D and Rigol M 2017, \href{https://journals.aps.org/prx/abstract/10.1103/PhysRevX.7.021012}{{\em Phys. Rev. X\/} {\bf 7} 021012}.

\bibitem{vidmar_xu_17}
Vidmar L, Xu W and Rigol M 2017, \href{https://link.aps.org/doi/10.1103/PhysRevA.96.013608}
{ {\em Phys. Rev. A} {\bf 96} 013608}.

\bibitem{zhang_vidmar_19}
Zhang Y, Vidmar L and Rigol M 2019, \href{https://link.aps.org/doi/10.1103/PhysRevA.99.063605}
{ {\em Phys. Rev. A} {\bf 99} 063605}.

\bibitem{zhang_vidmar_21}
Zhang Y, Vidmar L and Rigol M 2021, \href{https://link.aps.org/doi/10.1103/PhysRevA.104.L031303}
{ {\em Phys. Rev. A} {\bf 104} L031303}.

\bibitem{rigol2005fermionization}
Rigol M and Muramatsu A 2005, \href{https://journals.aps.org/prl/abstract/10.1103/PhysRevLett.94.240403}{ {\em Phys.~Rev.~Lett.} {\bf 94} 240403}.

\bibitem{sutherland_98}
Sutherland B 1998, \href{http://link.aps.org/doi/10.1103/PhysRevLett.80.3678}{ {\em Phys. Rev. Lett.} {\bf 80} 3678}.

\bibitem{minguzzi05}
Minguzzi A and Gangardt D~M 2005, \href{https://journals.aps.org/prl/abstract/10.1103/PhysRevLett.94.240404}{{\em Phys. Rev. Lett.} {\bf 94} 240404}.

\bibitem{bolech12}
Bolech C~J, Heidrich-Meisner F, Langer S, McCulloch I~P, Orso G and Rigol M
  2012, \href{https://journals.aps.org/prl/abstract/10.1103/PhysRevLett.109.110602}{ {\em Phys. Rev. Lett.} {\bf 109} 110602}.

\bibitem{campbell15}
Campbell A~S, Gangardt D~M and Kheruntsyan K~V 2015, \href{http://link.aps.org/doi/10.1103/PhysRevLett.114.125302}{ {\em Phys. Rev. Lett.}
  {\bf 114} 125302}.


\bibitem{mei16}
Mei Z, Vidmar L, Heidrich-Meisner F and Bolech C~J 2016, \href{http://link.aps.org/doi/10.1103/PhysRevA.93.021607}{ {\em Phys. Rev. A}
  {\bf 93} 021607(R)}.


\bibitem{kinoshita04}
Kinoshita T, Wenger T and Weiss D~S 2004, \href{https://www.science.org/doi/10.1126/science.1100700}{{\em Science} {\bf 305} 1125}.

\bibitem{paredes04}
Paredes B, Widera A, Murg V, Mandel O, F\"olling S, Cirac I, Shlyapnikov G,
  H\"ansch T~H and Bloch I 2004, \href{https://www.nature.com/articles/nature02530}{ {\em Nature} {\bf 429} 277}.

\bibitem{ronzheimer13}
Ronzheimer J~P, Schreiber M, Braun S, Hodgman S~S, Langer S, McCulloch I~P,
  Heidrich-Meisner F, Bloch I and Schneider U 2013, \href{https://journals.aps.org/prl/abstract/10.1103/PhysRevLett.110.205301}{ {\em Phys. Rev. Lett.}
  {\bf 110} 205301}.

\bibitem{vidmar_ronzheimer_15}
Vidmar L, Ronzheimer J~P, Schreiber M, Braun S, Hodgman S~S, Langer S,
  Heidrich-Meisner F, Bloch I and Schneider U 2015, \href{https://link.aps.org/doi/10.1103/PhysRevLett.115.175301}{ {\em Phys. Rev. Lett.}
  {\bf 115} 175301}.


\bibitem{wilson2020observation}
Wilson J~M, Malvania N, Le Y, Zhang Y, Rigol M and Weiss D~S 2020, \href{https://www.science.org/doi/10.1126/science.aaz0242}{ {\em
  Science\/} {\bf 367} 1461--1464}.

\bibitem{Penrose-Onsager56}
Penrose O and Onsager L 1956, \href{https://link.aps.org/doi/10.1103/PhysRev.104.576}{ {\em Phys. Rev.} {\bf 104} 576--584}.

\bibitem{pezer2007momentum}
Pezer R and Buljan H 2007, \href{https://journals.aps.org/prl/abstract/10.1103/PhysRevLett.98.240403}{ {\em Phys.~Rev.~Lett.} {\bf 98} 240403}.

\bibitem{forrester2003finite}
Forrester P, Frankel N, Garoni T and Witte N 2003, \href{https://journals.aps.org/pra/abstract/10.1103/PhysRevA.67.043607}{ {\em Phys.~Rev.~A} {\bf 67}
  043607}.

\bibitem{colcelli2018universal}
Colcelli A, Viti J, Mussardo G and Trombettoni A 2018, \href{https://journals.aps.org/pra/abstract/10.1103/PhysRevA.98.063633}{ {\em Phys.~Rev.~A} {\bf
  98} 063633}.

\bibitem{brun2017one}
Brun Y and Dubail J 2017, \href{https://scipost.org/SciPostPhys.2.2.012}{ {\em SciPost Phys.} {\bf 2} 012}.

\bibitem{brun2018inhomogeneous}
Brun Y and Dubail J 2018, \href{https://scipost.org/10.21468/SciPostPhys.4.6.037}{ {\em SciPost Phys.} {\bf 4} 037}.

\bibitem{scopa2020one}
Scopa S, Piroli L and Calabrese P 2020, \href{https://iopscience.iop.org/article/10.1088/1742-5468/abaed1/meta}{ {\em J. Stat. Mech.\/} {\bf 2020} 093103}.

\bibitem{Scopa_2023}
Scopa S, Ruggiero P, Calabrese P and Dubail J 2023, \href{https://doi.org/10.1103%2Fphysreva.108.013324}{ {\em Phys. Rev. A}
  {\bf 108} }.

\bibitem{ruggiero2021quantum}
Ruggiero P, Calabrese P, Doyon B and Dubail J 2021, \href{https://iopscience.iop.org/article/10.1088/1751-8121/ac3d68}{ {\em J. Phys. A: Math. Theor.} {\bf 55} 024003}.

\bibitem{ruggiero2020quantum}
Ruggiero P, Calabrese P, Doyon B and Dubail J 2020, \href{https://journals.aps.org/prl/abstract/10.1103/PhysRevLett.124.140603}{ {\em Phys. Rev. Lett.}
  {\bf 124} 140603}.

\bibitem{scopa2021exact}
Scopa S, Krajenbrink A, Calabrese P and Dubail J 2021, \href{https://iopscience.iop.org/article/10.1088/1742-5468/ac85eb/meta}{ {\em J.~Phys.~A:
  Math.~Theor.} {\bf 54} 404002}.

\bibitem{ruggiero2019conformal}
Ruggiero P, Brun Y and Dubail J 2019, \href{https://www.scipost.org/SciPostPhys.6.4.051/pdf}{ {\em SciPost Phys.} {\bf 6} 51}.

\bibitem{Collura2020domwallmelting}
Collura M, De~Luca A, Calabrese P and Dubail J 2020, \href{https://link.aps.org/doi/10.1103/PhysRevB.102.180409}{ {\em Phys. Rev. B} {\bf
  102} 180409}.


\bibitem{ares2022entanglement}
Ares F, Scopa S and Wald S 2022, \href{https://iopscience.iop.org/article/10.1088/1751-8121/ac8209/meta}{ {\em J. Phys. A: Math. Theor.} {\bf 55} 375301}.

\bibitem{scopa2022exact}
Scopa S, Calabrese P and Dubail J 2022, \href{https://scipost.org/SciPostPhys.12.6.207}{ {\em SciPost Phys.} {\bf 12} 207}.

\bibitem{Rottoli_2022}
Rottoli F, Scopa S and Calabrese P 2022, \href{http://dx.doi.org/10.1088/1742-5468/ac72a1}{ {\em J. Stat. Mech.} {\bf(2022)} 063103}.


\bibitem{Scopa2023domainwall}
Scopa S and Karevski D 2023, \href{http://dx.doi.org/10.1140/epjs/s11734-023-00845-1} {{\em Eur. Phys. J. Spec.
  Top.} {\bf 232} 1763}.


\bibitem{Capizzi_2023}
Capizzi L, Vanoni C, Calabrese P and Gambassi A 2023, \href{http://dx.doi.org/10.1088/1742-5468/ace3b4}{ {\em J. Stat. Mech.} {\bf (2023)} 073104}.

\bibitem{allegra2016inhomogeneous}
Allegra N, Dubail J, St{\'e}phan J~M and Viti J 2016, \href{https://iopscience.iop.org/article/10.1088/1742-5468/2016/05/053108}{ {\em J. Stat. Mech.} {\bf (2016)} 053108}.

\bibitem{Karevski2002}
Karevski D 2002, \href{http://dx.doi.org/10.1140/epjb/e20020139}{ {\em Eur. Phys. J. B } {\bf
  27} 147}.


\bibitem{olshanii2003short}
Olshanii M and Dunjko V 2003, \href{https://journals.aps.org/prl/abstract/10.1103/PhysRevLett.91.090401}{ {\em Phys. Rev. Lett.} {\bf 91} 090401}.

\bibitem{Eisler_2021}
Eisler V 2021, \href{https://dx.doi.org/10.1088/1751-8121/ac21e4}{ {\em J. Phys. A: Math. Theor.} {\bf
  54} 424002}.

\bibitem{fang2016momentum}
Fang B, Johnson A, Roscilde T and Bouchoule I 2016, \href{https://journals.aps.org/prl/abstract/10.1103/PhysRevLett.116.050402}{ {\em Phys.~Rev.~Lett.}
  {\bf 116} 050402}.

\bibitem{schemmer2019generalized}
Schemmer M, Bouchoule I, Doyon B and Dubail J 2019, \href{https://journals.aps.org/prl/abstract/10.1103/PhysRevLett.122.090601}{ {\em Phys.~Rev.~Lett.}
  {\bf 122} 090601}.

\bibitem{bouchoule2022generalized}
Bouchoule I and Dubail J 2022, \href{https://iopscience.iop.org/article/10.1088/1742-5468/ac3659}{ {\em J.~Stat.~Mech.} {\bf (2022)} 014003}.

\bibitem{antal1999transport}
Antal T, R\'acz Z, R\'akos A and Sch\"utz G~M 1999, \href{https://link.aps.org/doi/10.1103/PhysRevE.59.4912}{ {\em Phys. Rev. E} {\bf
  59} 4912}.


\bibitem{antal2008log}
Antal T, Krapivsky P~L and R\'akos A 2008, \href{https://link.aps.org/doi/10.1103/PhysRevE.78.061115}{ {\em Phys. Rev. E} {\bf 78}
  061115}.

\bibitem{Platini_2007}
Platini T and Karevski D 2007, \href{https://dx.doi.org/10.1088/1751-8113/40/8/002}{ {\em J. Phys. A: Math. Theor.} {\bf 40} 1711}.


\bibitem{gerbier2003momentum}
Gerbier F, Thywissen J~H, Richard S, Hugbart M, Bouyer P and Aspect A 2003, \href{https://journals.aps.org/pra/abstract/10.1103/PhysRevA.70.013607}{ {\em
  Phys.~Rev.~A} {\bf 67} 051602}.

\bibitem{regal2005momentum}
Regal C, Greiner M, Giorgini S, Holland M and Jin D 2005, \href{https://journals.aps.org/prl/abstract/10.1103/PhysRevLett.95.250404}{ {\em
  Phys.~Rev.~Lett.} {\bf 95} 250404}.

\bibitem{jukic2009momentum}
Juki{\'c} D, Klajn B and Buljan H 2009, \href{https://journals.aps.org/pra/abstract/10.1103/PhysRevA.79.033612}{ {\em Phys. Rev. A} {\bf 79} 033612}.

\bibitem{vaidya1979one}
Vaidya H~G and Tracy C 1979, \href{https://pubs.aip.org/aip/jmp/article-abstract/20/11/2291/449636/One-particle-reduced-density-matrix-of?redirectedFrom=fulltext}{ {\em J.~Math.~Phys.} {\bf 20} 2291--2312}.

\bibitem{lenard1964momentum}
Lenard A 1964, \href{https://pubs.aip.org/aip/jmp/article-abstract/5/7/930/378564/Momentum-Distribution-in-the-Ground-State-of-the?redirectedFrom=fulltext}{ {\em J.~Math.~Phys.} {\bf 5} 930--943}.

\bibitem{lenard1972some}
Lenard A 1972, \href{https://projecteuclid.org/journals/pacific-journal-of-mathematics/volume-42/issue-1/Some-remarks-on-large-Toeplitz-determinants/pjm/1102968015.full}{ {\em Pac.~J.~Math.} {\bf 42} 137--145}.


\bibitem{bertini2016transport}
Bertini B, Collura M, De~Nardis J and Fagotti M 2016, \href{https://journals.aps.org/prl/abstract/10.1103/PhysRevLett.117.207201}{ {\em Phys.~Rev.~Lett.}
  {\bf 117} 207201}.

\bibitem{castro2016emergent}
Castro-Alvaredo O~A, Doyon B and Yoshimura T 2016, \href{https://journals.aps.org/prx/abstract/10.1103/PhysRevX.6.041065}{ {\em Phys.~Rev.~X} {\bf 6}
  041065}.

\bibitem{Collura_2018}
Collura M, De~Luca A and Viti J 2018, \href{http://dx.doi.org/10.1103/PhysRevB.97.081111}{ {\em Physical Review B\/} {\bf 97} 081111 (R)}.

\bibitem{PhysRevA.89.063603}
Halimeh J~C, W\"ollert A, McCulloch I, Schollw\"ock U and Barthel T 2014, \href{https://link.aps.org/doi/10.1103/PhysRevA.89.063603}{ {\em
  Phys. Rev. A} {\bf 89} 063603}.


\bibitem{Capizzi_2023b}
Capizzi L, Scopa S, Rottoli F and Calabrese P 2023, \href{https://dx.doi.org/10.1209/0295-5075/acb50a}{ {\em Europhysics Letters\/}
  {\bf 141} 31002}.

\bibitem{PhysRevB.99.174403}
Gruber M and Eisler V 2019, \href {https://link.aps.org/doi/10.1103/PhysRevB.99.174403}{{\em Phys. Rev. B\/} {\bf 99} 174403}.

\bibitem{dubail2017conformal}
Dubail J, St{\'e}phan J~M, Viti J and Calabrese P 2017, \href{https://scipost.org/SciPostPhys.2.1.002}{ {\em SciPost Phys.}
  {\bf 2} 002}.

\bibitem{fagotti2017higher}
Fagotti M 2017, \href{https://journals.aps.org/prb/abstract/10.1103/PhysRevB.96.220302}{ {\em Phys. Rev. B} {\bf 96} 220302}.

\bibitem{fagotti2020locally}
Fagotti M 2020, \href{https://scipost.org/SciPostPhys.8.3.048}{ {\em SciPost Phys.} {\bf 8} 048}.

\bibitem{doyon2017large}
Doyon B, Dubail J, Konik R and Yoshimura T 2017, \href{https://journals.aps.org/prl/abstract/10.1103/PhysRevLett.119.195301}{ {\em Phys. Rev. Lett.} {\bf
  119} 195301}.

\bibitem{hinarejos2012wigner}
Hinarejos M, P{\'e}rez A and Ba{\~n}uls M~C 2012, \href{https://iopscience.iop.org/article/10.1088/1367-2630/14/10/103009/meta}{ {\em New J. Phys.}
  {\bf 14} 103009}.
  
  \bibitem{wendenbaum2013hydro}
Wendenbaum P, Collura M and Karevski D 2013, \href{https://link.aps.org/doi/10.1103/PhysRevA.87.023624}{ {\em Phys. Rev. A} {\bf 87}
  023624 }.


\bibitem{shashi2011nonuniversal}
Shashi A, Glazman L~I, Caux J~S and Imambekov A 2011, \href{https://journals.aps.org/prb/abstract/10.1103/PhysRevB.84.045408}{{\em Phys. Rev. B} {\bf
  84} 045408}.

\bibitem{shashi2012exact}
Shashi A, Panfil M, Caux J~S and Imambekov A 2012, \href{https://journals.aps.org/prb/abstract/10.1103/PhysRevB.85.155136}{ {\em Phys. Rev. B} {\bf 85}
  155136}.

\bibitem{bondesan2015chiral}
Bondesan R, Dubail J, Faribault A and Ikhlef Y 2015, \href{https://iopscience.iop.org/article/10.1088/1751-8113/48/6/065205}{ {\em J. Phys. A: Math. Theor. {\bf 48} 065205}.}

\bibitem{eliens2017quantum}
Eli{\"e}ns S 2017, \href{https://pure.uva.nl/ws/files/9247453/OnQuantumSeas_Eliens_proefschrift_complete.pdf}{ {\em On quantum seas}}, Ph.D. thesis ~--~ Univ. of Amsterdam.

\bibitem{bouchoule2021breakdown}
Bouchoule I and Dubail J 2021, \href{https://journals.aps.org/prl/abstract/10.1103/PhysRevLett.126.160603}{ {\em Phys. Rev. Lett.} {\bf 126} 160603}

\bibitem{Yukalov_2005}
Yukalov V~I and Girardeau M~D 2005, \href{https://dx.doi.org/10.1002/lapl.200510011}{ {\em Laser Phys. Lett.} {\bf 2} 375}.

\bibitem{collura2013equilibration}
Collura M, Sotiriadis S and Calabrese P 2013, \href{https://journals.aps.org/prl/abstract/10.1103/PhysRevLett.110.245301}{{\em Phys. Rev. Lett.}
  {\bf 110} 245301}.

\bibitem{collura2013quench}
Collura M, Sotiriadis S and Calabrese P 2013, \href{https://iopscience.iop.org/article/10.1088/1742-5468/2013/09/P09025/meta}{ {\em J. Stat. Mech.} {\bf( 2013)} P09025}.


\end{thebibliography}

\end{document}